\documentclass[letterpaper,twocolumn,10pt]{article}
\usepackage{usenix,epsfig,endnotes}
\usepackage[available]{usenixbadges}

\usepackage{xcolor}
\usepackage{tikz}
\usepackage{amsmath}
\usepackage{textcomp}
\usepackage{multirow}
\usepackage{graphicx}
\usepackage[most]{tcolorbox}
\usepackage{algorithm,algpseudocode}
\usepackage{optidef}
\usepackage{threeparttable}
\usepackage{makecell}
\usepackage{arydshln}
\usepackage{xspace}
\usepackage{tabularray}
\usepackage{amssymb} 
\usepackage{url} 

\newcommand{\sysname}{{AFTUNE}\xspace}
\newcommand{\be}{\begin{equation}}
\newcommand{\ee}{\end{equation}}

\begin{document}

\title {Trusting What You Cannot See: Auditable Fine-Tuning and Inference for Proprietary AI}

\author{
{\rm Heng Jin}\\
Virginia Tech
\and
{\rm Chaoyu Zhang}\\
Virginia Tech
\and
{\rm Hexuan Yu}\\
Virginia Tech
\and
{\rm Shanghao Shi}\\
Washington University in St.\ Louis
\and
{\rm Ning Zhang}\\
Washington University in St.\ Louis
\and
{\rm Y.\ Thomas Hou}\\
Virginia Tech
\and
{\rm Wenjing Lou}\\
Virginia Tech
}

\maketitle

\subsection*{Abstract}
Cloud-based infrastructure has become the dominant platform for deploying large models, particularly large language models (LLMs). Fine-tuning and inference are increasingly delegated to cloud providers for simplified deployment and access to proprietary models, yet this creates a fundamental trust gap. Although cryptographic and TEE-based verification approaches exist, prohibitive proving costs and limited TEE memory prevent them from scaling to modern LLMs, leaving clients unable to practically audit these processes. This lack of transparency creates concrete security risks that can silently compromise service integrity. 

We present \sysname, an auditable and verifiable framework that ensures the computational integrity of cloud-based fine-tuning and inference. \sysname incorporates a lightweight recording and spot-check mechanism that produces verifiable traces of execution. These traces enable clients to later audit whether the fine-tuning and inference processes followed the agreed configurations, by verifying sampled execution blocks inside a TEE, each covering only a small portion of the model and the execution trace. Our evaluation shows that \sysname adds modest overhead and makes auditing practical for clients.

\section{Introduction}
\label{sec:introduction}
Cloud-based infrastructures have become the dominant platform for deploying and customizing modern AI models~\cite{openai_model_optimization_2025, anthropic_fine_tune_claude3_haiku_2024, googlecloud_gemini_supervised_tuning_2025, amazon_aws_bedrock_2025, oracle_host_llms_nvidia_oci_2025}. Infrastructure providers now host LLMs and provide managed APIs for fine-tuning, inference, and Model-as-a-Service deployments. For instance, OpenAI~\cite{openai_model_optimization_2025} offers GPT fine-tuning and hosted inference through its API, where clients submit data and access tuned models exclusively via API endpoints. Hundreds of thousands of models have been fine-tuned using OpenAI's service~\cite{openai2024finetuning}. As models grow ever larger, enterprises are increasingly inclined to use the cloud to host fine-tuning and inference even for open-source models, rather than maintain local infrastructure.

However, this division of responsibility creates a fundamental visibility gap. Clients provide data and configurations but cannot independently verify whether providers execute fine-tuning and inference as contracted, and lack evidence linking delivered models or inference outputs to actual internal computation. Untrusted providers may downgrade the service, embed biases, or even implant backdoors into the model. Appendix~\ref{app:attacks} outlines possible attacks along with attacker objectives and concrete examples. At the same time, service providers lack the means to prove that they executed fine-tuning and inference as contracted, especially when they cannot provide model weights to the client (e.g., under API-only or proprietary-model deployments).

For proprietary models such as GPT and Gemini, parameter confidentiality prevents clients from directly inspecting models and forces both fine-tuning and inference into the cloud. Protecting only one process is insufficient, as a provider that faithfully executes fine-tuning may still misbehave during inference, and vice versa. This necessitates a unified verification framework that protects both processes.

\vspace{3pt} \noindent \textbf{Existing literature and challenges on verifying fine-tuning and inference of LLMs.} Verifiable computation in cloud environments has been a long-standing research topic. Most prior works aiming to enable verification without exposing details fall into two categories, zero-knowledge proof (ZKP) based approaches and trusted execution environment (TEE) based approaches. However, the increasing scale of modern large models introduces new challenges that existing approaches fail to address, due to the enormous memory footprint and computational requirements of these models.

ZKP-based approaches~\cite{zkgpt,sun2024zkllm,xie2025zkpytorch,hao2024scalable,abbaszadeh2024zero, Liao2025VeriLoRA} provide mathematical integrity for computations but incur prohibitive overhead due to the complexity of generating proofs for large-scale neural network operations. For example, state-of-the-art ZKP systems for model inference~\cite{sun2024zkllm,zkgpt} incur commitment overheads on the order of thousands of times the cost of standard plaintext execution (e.g., up to 15 minutes of overhead for a single-token inference of a 13B model~\cite{sun2024zkllm}). VeriLoRA~\cite{Liao2025VeriLoRA} applies ZKP to LLM training, but the same overhead barrier restricts it to LoRA, where only a small portion of parameters is trained. Among TEE-based approaches, \cite{Steiakakis2026TEE, Tan2025PipeLLM} run the inference pipeline within TEEs, while \cite{secureTF, mo2021ppfl, Dhar25Guardain} place the entire pipeline inside them. Whether using TEEs on CPUs, GPUs, or NPUs, those approaches require the TEE to hold the entire model.

\vspace{3pt} \noindent \textbf{Our work.} Given the prohibitive proof generation overhead inherent to ZKP-based approaches, we adopt the TEE-based paradigm. However, to make TEE-based verification practical for large language models, the fundamental challenges mentioned above must be resolved. We present \sysname, a framework that enables verifiable integrity for cloud-based fine-tuning and inference by addressing the following challenges:

\emph{Challenge 1) Memory constraints:} Modern LLMs exceed the trusted memory available to a single TEE instance and are routinely deployed via sharding across multiple GPUs or nodes. Fine-tuning makes this worse, as training requires additional persistent state such as optimizer moments and variance, plus large temporary memory for activations, gradients, and batch buffers. Directly applying existing run-inside-TEE techniques to fine-tuning is therefore impractical.

\begin{figure}[ht]
    \centering
    \includegraphics[width=0.95\linewidth]{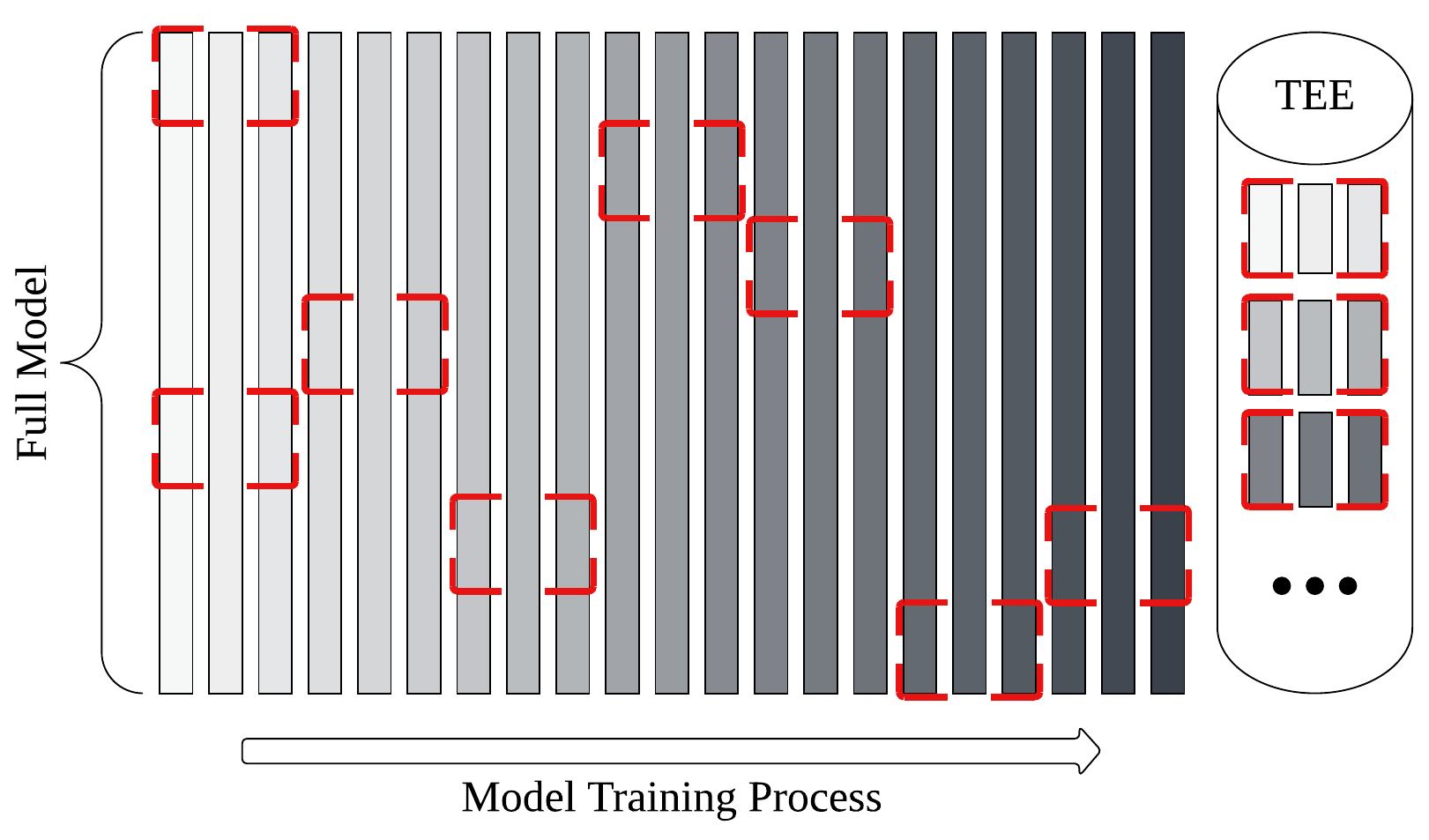}
    \caption{Each rectangle represents the full model at one training step. The red dashed border indicates blocks that will be independently verified, each of which contains only a portion of the training procedure.}
    \label{fig:model_partition}
  \end{figure}

\emph{Solution 1) Block decomposition with compositional verification:} We exploit a key structural property of neural computation. Within a contiguous region of layers and steps, all internal states are deterministic functions of the region's boundary states. This allows us to decompose execution into independently verifiable blocks along layer and step dimensions, recording only boundary states, as illustrated in Figure~\ref{fig:model_partition}. Each block can be replayed inside an attested TEE as an atomic unit, and enforcing boundary continuity makes these local checks compose into end-to-end verification without requiring the full model or trace to reside in trusted memory.

\emph{Challenge 2) Cost of evidence generation and verification:} Verifying the provider introduces two unavoidable costs. The provider must produce binding evidence during normal execution by computing commitments over large boundary tensors, and since clients cannot access proprietary model weights, verification must be performed on the provider side inside a TEE. At LLM scale, both are expensive. Naive cryptographic hashing over millions of tensor elements can bottleneck throughput, and recomputation-based audits become costly when replaying long segments. This creates a tension between integrity, which demands dense evidence and thorough checking, and practicality, which requires low execution overhead and affordable verification costs.

\emph{Solution 2) Map-reduce style hashing and sampling-based verification:} We address execution overhead by making commitments accelerator-friendly. Instead of sequentially hashing each large tensor, we partition tensors into chunks, hash chunks in parallel on the accelerator, and aggregate them into a single commitment. We address verification overhead by sampling, where the client verifies only a sampled subset of blocks with randomness hidden until audit time, yielding probabilistic detection guarantees that compound over repeated audits. Because block checks are independent, audits can be parallelized across TEEs without consuming accelerator resources.

The main contributions of this paper are:
\begin{itemize}
    \item We propose \sysname, a verifiable framework for cloud-based fine-tuning and inference that decouples verification from the fine-tuning process through block-based evidence generation. This allows \sysname to overcome the resource constraints of the TEE, enabling clients to verify the fine-tuning process while preserving the confidentiality of proprietary model parameters.
    \item We introduce a sampling-based verification strategy that provides probabilistic detection guarantees at practical costs, along with a map-reduce style hashing scheme for efficient commitment generation.
    \item We implement \sysname and integrate it into CUDA-based training and inference pipelines. We evaluate our system on real TEE hardware with multiple open-source models across different model families and sizes, using both full fine-tuning and parameter-efficient fine-tuning. The results show that \sysname incurs low training time overhead while keeping verification costs modest.
\end{itemize}
\section{Related Work}
\label{sec:background}

Various approaches have been proposed to establish trust in cloud-based AI services, spanning behavioral testing, cryptographic verification, and hardware-based isolation.

Red-team evaluations can expose functional or safety issues in deployed models~\cite{feffer2024red, lin2025against}, but only test known threat patterns and cannot verify whether fine-tuning followed the contracted procedure or whether the served model corresponds to the expected checkpoint. These approaches provide behavioral validation rather than process integrity guarantees. 
Cryptographic approaches, particularly ZKP, offer strong integrity guarantees for computation~\cite{zkgpt,riasi2025zero,sun2024zkllm,xie2025zkpytorch,hao2024scalable,abbaszadeh2024zero,Liao2025VeriLoRA}. As discussed in Section~\ref{sec:introduction}, existing ZKP-based approaches incur prohibitive computational overhead, making them impractical for production deployment.

Trusted execution environments provide hardware-enforced isolation that preserves code and data integrity even when the OS is compromised. Common TEEs include ARM TrustZone~\cite{arm_trustzone}, Intel SGX~\cite{intel_sgx}, and RISC-V Keystone~\cite{riscv_keystone}. TEEs support remote attestation, allowing verifiers to cryptographically validate software state through hardware-signed measurements. Prior work has applied TEE-backed remote attestation to model execution~\cite{chen2019deepattest,mo2021ppfl,zhao2021sear,zhang2020enabling,Steiakakis2026TEE,secureTF, Tan2025PipeLLM, Dhar25Guardain}.
%, following state-of-the-art attestation frameworks~\cite{nunes2020apexRA1, abera2016cRA2, abera2019diatRA3, sun2020oatRA4, surminski2021realswattRA5, zhang2021recfaRA6, wang2023ariRA7, ammar2024bridgingRA8}. 
However, as discussed in Section~\ref{sec:introduction}, running entire fine-tuning or inference pipelines inside TEEs is impractical for large models due to memory constraints and performance overhead.

The commit-and-sample paradigm has long been used to audit outsourced computation. Prior works used probabilistic spot-checking to detect whether untrusted remote hosts skipped assigned tasks or fabricated outputs~\cite{Monrose1999DistributedEW}. The same idea later appeared in volunteer computing, where tasks with known answers reduced worker error rates at low overhead~\cite{Sarmenta2002}, in fair payment protocols that compensated workers only after verified completion~\cite{Carbunar2010, Chen2012}, and in chaff injection schemes that exposed falsified results~\cite{DuGoodrich2005}. \sysname brings this paradigm to machine learning fine-tuning and inference, where new challenges arise, such as computation at tensor scale and the layer-wise execution flow of neural networks.

Some recent attempts address verification through indirect methods, but their scope remains limited. SVIP~\cite{sun2024svip} requires service providers to return hidden states and verifies model identity through designed tasks, but only addresses the problem of providers using cheaper models during inference. VTUNE~\cite{zhang2024vtune} inserts backdoors into training datasets and detects their presence afterward, but only catches cases where the provider completely skips fine-tuning and uses the base model.
\section{Threat Model}
\label{sec:threat-model}

We assume the client supplies the training dataset and specifies fine-tuning hyperparameters, while the service provider operates the training infrastructure. We consider proprietary model fine-tuning, where the client cannot access model parameters at any stage and must rely entirely on the provider for both training and inference.

We assume the service provider publicly releases the hash commitment of the base model as the initial trusted checkpoint for verifiable fine-tuning. We assume the cloud platform provides TEE-enabled infrastructure supporting remote attestation, allowing the client to verify TEE measurements and establish secure channels to provision verification workloads to attested enclaves. Due to TEE memory constraints, we do not assume that the entire model can be loaded into the TEE at once for either training or inference. We do assume, however, that no single parameter tensor (i.e., an individual weight array that is the unit of computation in the model) exceeds the TEE's memory capacity. In our TEE setting, we have not observed any existing model, regardless of parameter count, that fails to satisfy this requirement.

We assume the attacker is a misbehaving or compromised cloud provider who aims to manipulate training or inference processes while evading detection. The adversary may control all software and infrastructure outside the TEE, fully understand \sysname's design and verification mechanisms, and perform training manipulation or inference attacks as detailed in Appendix~\ref{app:attacks}. While potential attacks against TEEs or cryptographic hash functions exist, addressing or mitigating such attacks is outside the scope of this work.

We assume the defender is the client of the cloud service provider. The objective of \sysname is to enable the client to verify that training adhered to the contracted dataset, model, and hyperparameters, and that inference outputs are genuine predictions from the verified model, without requiring access to model parameters. % or continuous monitoring. 
\section{Overview} 
\label{sec:overview}

\begin{figure*}[t]
\centering
\includegraphics[width=0.85\linewidth]{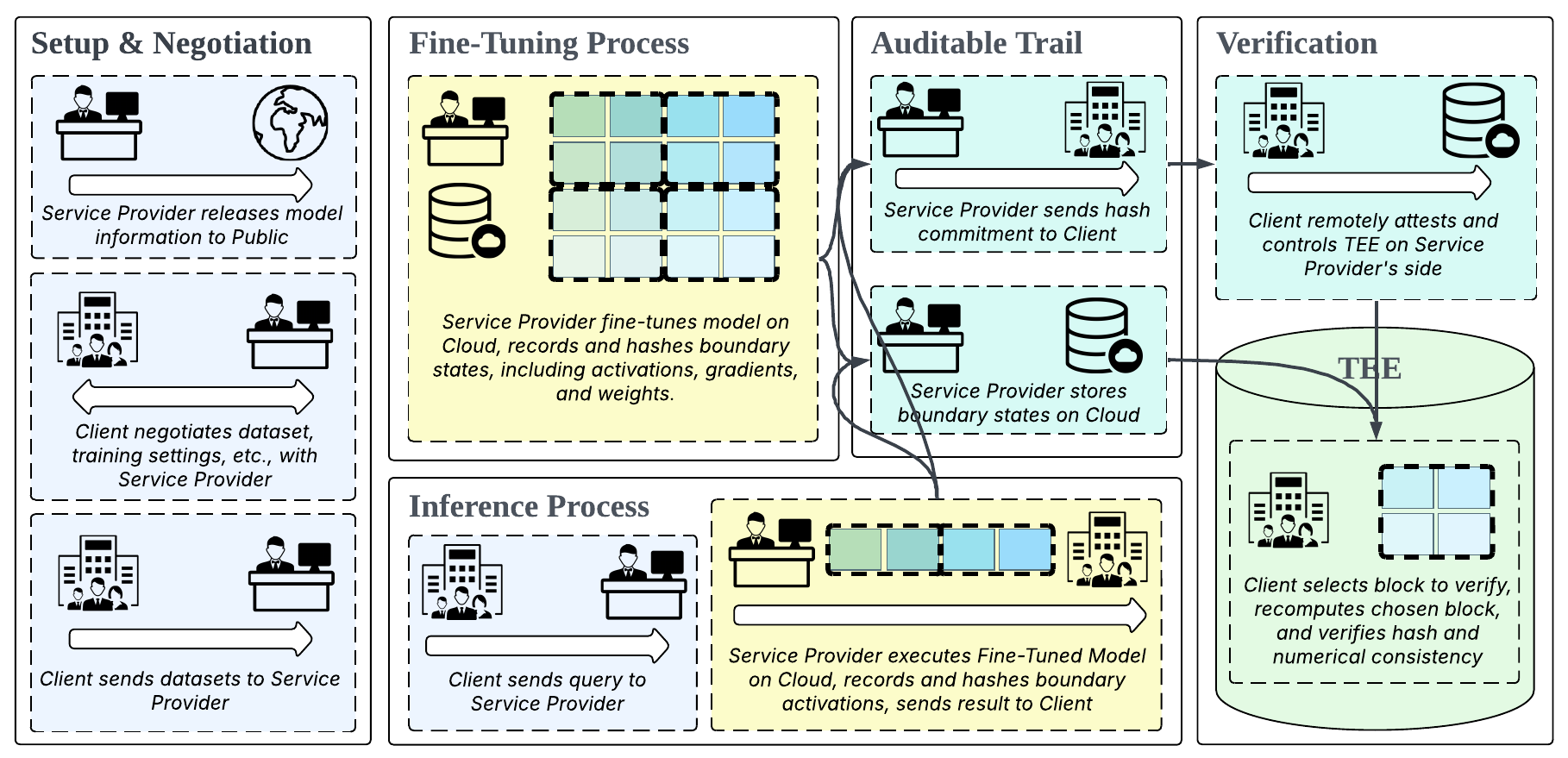}
\caption{
\textbf{\sysname workflow overview.} The system operates through setup and negotiation, fine-tuning, and inference phases. During execution, the provider commits boundary state hashes to the client and stores boundary states in cloud storage. Clients can verify training or inference computations on demand through TEE-based selective recomputation of sampled regions.
}
\label{fig:aftune-workflow}
\end{figure*}

\begin{figure}[ht]
    \centering
    \includegraphics[width=0.95\linewidth]{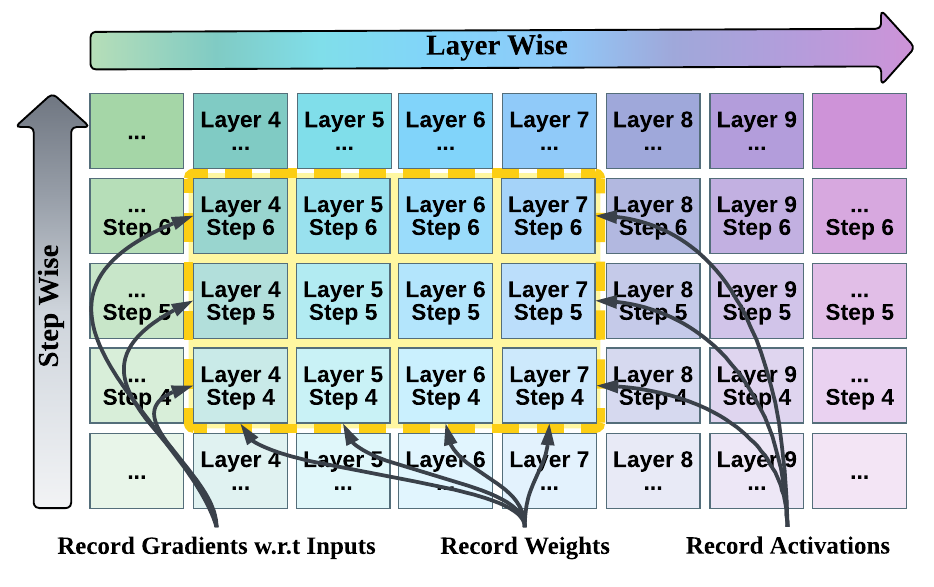}
    \caption{Two-dimensional block structure organizing the training procedure. The dashed border shows a single block spanning 4 layers and 3 steps. \sysname records only boundary states: activations and gradients at layer block edges, and parameters and optimizer states at step block boundaries. Each block serves as an atomic, independently verifiable unit.}
    \label{fig:block_structure}
  \end{figure}

\subsection{Design Rationale}
\label{subsec:approach}
TEEs have seen significant growth over the past decade, with GPU TEEs now emerging. However, due to the size of modern large models, it remains impractical to train them on a single node, let alone within the TEE on that node. TEEs inherently limit computational capability due to security constraints~\cite{Lee2025GPUTEE}. Furthermore, modern large model training uses heterogeneous accelerators such as GPUs, NPUs, FPGAs, and ASICs to tackle performance challenges~\cite{Kachris2025acce,Silvano2025acce,Emani2023aace,Yan2025acce,Tang2025aace}. While CPU TEEs are widely available, TEE support for these accelerators remains inconsistent, and extending TEE-based verification across such distributed heterogeneous fabrics presents unique technical and deployment challenges. In this work, we take a different approach. Rather than requiring TEEs on all components of the training infrastructure, we anchor trust in widely available CPU TEEs and extend this trust to the fine-tuning process without requiring systematic changes to the complex training infrastructure.

Our approach relies on the practical availability of TEE infrastructure, supported by the growing adoption of confidential computing across major cloud platforms such as Google Cloud~\cite{googleConfidentialComputing} and AWS~\cite{amazonLargeLanguage}. We leverage remote TEE enclaves on the provider side. Clients specify verification tasks that execute inside the TEE, where the enclave accesses model parameters, verifies computations, and returns only verification results to the client. This design ensures model parameters never leave the secure enclave while still providing clients with attestation of correct execution. Instead of executing the entire training process inside a TEE, \sysname decouples training execution from verification. The key insight is that we do not need real-time protection of computations during training. Instead, we generate cryptographic evidence during execution and verify correctness afterward. This allows the primary fine-tuning workload to run at full speed on standard accelerator infrastructure while recording cryptographic commitments, using TEEs only for lightweight recomputation-based verification after execution completes. This design avoids the resource and performance constraints of full TEE encapsulation while still providing verifiability.

\subsection{System Workflow}
\label{subsec:overview_verification}
Verification in \sysname addresses two conceptually distinct properties. The first is stored data integrity, which concerns whether the recorded tensors, hashes, and checkpoints faithfully reflect what was produced during training. The second is computational correctness, which concerns whether training computations such as forward passes, backward passes, and weight updates were performed honestly. \sysname enforces the first by hash matching, confirming that stored data has not been tampered with. It then enforces the second by re-executing the computation in the TEE and comparing the results against the recorded outputs. Because stored data integrity is a precondition for computational correctness, both checks are unified into a single verification pipeline.

Figure~\ref{fig:aftune-workflow} illustrates \sysname's workflow. During training and inference, \sysname partitions the model layers into contiguous layer blocks and records boundary states at the layer-wise interfaces between adjacent blocks, then commits their hashes to the client, creating a cryptographic binding without requiring TEE involvement in the main computation path. The committed hashes may be signed with the service provider's private key to prove authenticity. Verification occurs on demand after execution completes. The client selects a target region (layer and step ranges for training, or layer ranges for inference) to verify. We refer to such a region as a block. A TEE instance recomputes only the selected block, then validates the recorded states through hash matching and verifies computational correctness by comparing recomputed values against recorded states within floating-point tolerance. Because recomputation targets small regions rather than full execution traces, TEE resource constraints are satisfied.

While our design focuses on proprietary model fine-tuning, it also applies to open-source model scenarios where clients outsource training but may obtain final weights for offline verification and self-hosted inference. Access to weights makes verification strictly easier, so our framework remains directly applicable.

\subsection{Block Structure and Trade-offs}
\label{subsec:overview_block}
To balance verification granularity with storage and computational efficiency, \sysname organizes the training procedure into a two-dimensional grid of blocks, as shown in Figure~\ref{fig:block_structure}. Specifically, \sysname groups consecutive layers into layer blocks and consecutive training steps into step blocks, creating a grid where each cell represents their intersection. Hashes and intermediate states are recorded only at block boundaries, significantly reducing hash computations and stored audit data compared to per-layer, per-step recording. Clients can verify any block independently, with recomputation cost scaling proportionally to block size. Hash computation itself is parallelized through a map-reduce scheme that partitions each tensor into chunks, computes hashes in parallel on the accelerator, and aggregates them into a single commitment. Since each block's commitment contains only hash values, the cost of transmitting and storing commitments is negligible. The layer block size governs a trade-off among multiple objectives. Larger blocks contain more layers, reducing the number of boundaries and thus the number of hash computations and the volume of intermediate states that must be stored. However, blocks cannot grow arbitrarily large, since the TEE must load a whole block during recomputation, and the block size is therefore bounded by the TEE's memory capacity. Smaller blocks also allow clients to achieve better detection probability under the same sampling rate, as we explain in Section~\ref{subsec:sampling_detection}.

\section{Design}
\label{sec:design}

\subsection{Notation and Problem Formulation}
\label{subsec:notation}

We formalize the fine-tuning verification problem and establish notation used throughout this section. Consider a neural network model with $L$ layers indexed by $\{l_0, l_1, \ldots, l_{L-1}\}$. Training proceeds for $T$ steps indexed by $\{t_0, t_1, \ldots, t_{T-1}\}$. For each layer $l$ and step $t$, we denote $x_{l,t} \in \mathbb{R}^{d_l}$ and $y_{l,t} \in \mathbb{R}^{d_l'}$ as the input and output activations, respectively. The corresponding gradients are denoted $\nabla x_{l,t} \in \mathbb{R}^{d_l}$ (input gradient) and $\nabla y_{l,t} \in \mathbb{R}^{d_l'}$ (output gradient). Each layer $l$ has associated parameters $\theta_{l,t} \in \mathbb{R}^{p_l}$ and optimizer state $s_{l,t}$ (e.g., momentum and variance for AdamW) at step $t$. The complete model parameters at step $t$ are $\theta_t = \{\theta_{l,t}\}_{l=0}^{L-1}$ and the optimizer states are $s_t = \{s_{l,t}\}_{l=0}^{L-1}$.

The forward pass at layer $l$ and step $t$ computes $y_{l,t} = f_l(x_{l,t}; \theta_{l,t})$. The backward pass computes $\nabla x_{l,t} = \frac{\partial \mathcal{L}_t}{\partial x_{l,t}}$ and $\nabla \theta_{l,t} = \frac{\partial \mathcal{L}_t}{\partial \theta_{l,t}}$ via backpropagation, where $\mathcal{L}_t$ is the loss at step $t$. After each step, parameters are updated via an optimizer as $\theta_{l,t+1}, s_{l,t+1} = \text{Update}(\theta_{l,t}, s_{l,t}, \nabla \theta_{l,t}, \eta)$ for each layer $l$, where $\eta$ denotes hyperparameters such as the learning rate.

\textbf{Training verification.} Given an initial model with parameters $\theta_0$ whose cryptographic hash $h_0 = H(\theta_0)$ is known and trusted, a training dataset $\mathcal{D} = \{(\mathbf{x}_i, \mathbf{y}_i)\}_{i=1}^{|\mathcal{D}|}$, and a hyperparameter configuration $\mathcal{H}$, the client verifies that the provider's claimed checkpoint $\theta_T$ was produced by executing $T$ training steps according to $\mathcal{H}$ on dataset $\mathcal{D}$, starting from $\theta_0$. Furthermore, the client verifies that intermediate training states (activations, gradients) were computed correctly at each layer and step, and that no data poisoning, training degradation, or parameter tampering occurred.

\textbf{Inference verification.} Given a verified model with parameters $\theta^*$ (whose hash has been established through commitment and verification), an input $\mathbf{x}$, and the provider's claimed output $\mathbf{y}$, the client verifies that $\mathbf{y}$ was produced by computing the forward pass through all $L$ layers with parameters $\theta^*$ on input $\mathbf{x}$, and that no model substitution, output fabrication, or computation tampering occurred.

\subsection{Two-Dimensional Block Structure}
\label{subsec:block_structure}

A naive approach to verifiable training would record activations and gradients at every layer for every training step, incurring prohibitive storage costs that scale as $O(L \times T)$ for a model with $L$ layers trained for $T$ steps. For large models and long training runs, this quickly becomes impractical. Moreover, computing cryptographic commitments requires $L \times T$ hash operations, introducing significant computational overhead. A similar problem exists for inference, where per-layer recording would scale linearly with model depth.

The key observation is that not all intermediate states need explicit storage. Given activations at layer $l$ and model parameters, activations at layer $l+1$ can be deterministically recomputed via forward propagation. Similarly, given gradients at layer $l+1$, gradients at layer $l$ can be recomputed via backpropagation. This observation motivates partitioning layers and steps into blocks, recording states only at block boundaries, and reconstructing all interior states through recomputation during verification.

\sysname organizes the training procedure into a two-dimensional grid of blocks along the layer and step dimensions. Define the layer block size $B_L$ and step block size $B_S$. The $L$ layers are partitioned into $\lceil L / B_L \rceil$ \textit{layer blocks}:
\begin{equation}
\text{LB}_i = \{l_{i \cdot B_L}, l_{i \cdot B_L + 1}, \ldots, l_{\min((i+1) \cdot B_L - 1, L-1)}\}
\end{equation}
for $i = 0, 1, \ldots, \lceil L / B_L \rceil - 1$. Similarly, the $T$ training steps are partitioned into $\lceil T / B_S \rceil$ \textit{step blocks}:
\begin{equation}
\text{SB}_j = \{t_{j \cdot B_S}, t_{j \cdot B_S + 1}, \ldots, t_{\min((j+1) \cdot B_S - 1, T-1)}\}
\end{equation}
for $j = 0, 1, \ldots, \lceil T / B_S \rceil - 1$. Each combination of a \textit{layer block} and a \textit{step block} defines a \textit{block cell}:
\begin{equation}
\text{Block}_{i,j} = \text{LB}_i \times \text{SB}_j.
\end{equation}

For each $\text{Block}_{i,j}$, we record boundary activations $x_{i \cdot B_L, t}$ (input to first layer) and $y_{(i+1) \cdot B_L - 1, t}$ (output from last layer) for all steps $t \in \text{SB}_j$. Similarly, we record the corresponding gradients $\nabla x_{i \cdot B_L, t}$ and $\nabla y_{(i+1) \cdot B_L - 1, t}$ for all $t \in \text{SB}_j$. Additionally, parameters $\theta_{l,t}$ and optimizer states $s_{l,t}$ for layers $l \in \text{LB}_i$ are recorded at step block boundaries $t = j \cdot B_S$ and $t = (j+1) \cdot B_S$. 

The correctness of this boundary-only recording strategy relies on the continuity property of neural network computation, where the output of layer block $\text{LB}_i$ serves as the input to layer block $\text{LB}_{i+1}$, formally $y_{l_{\text{last}(i)}, t} = x_{l_{\text{first}(i+1)}, t}$ for all steps $t$. Similarly, during backpropagation, the input gradient to block $\text{LB}_{i+1}$ equals the output gradient of block $\text{LB}_i$: $\nabla x_{l_{\text{first}(i+1)}, t} = \nabla y_{l_{\text{last}(i)}, t}$. This continuity ensures that boundary states for adjacent blocks chain together, enabling complete verification of the forward and backward passes across all layers for any training step by recomputing within each block and validating boundary consistency.

Boundary continuity enables storage and computation deduplication. Since the output activation of block $\text{LB}_i$ is identical to the input activation of block $\text{LB}_{i+1}$, we need only store this tensor once and compute its cryptographic hash once, despite serving as a boundary for two adjacent blocks. The same principle holds for gradients during backpropagation, as shown in Figure~\ref{fig:block_structure}. As a result, instead of storing and hashing $2 \times \lceil L / B_L \rceil$ boundary states per step (an input and output for each block), we only store and hash $\lceil L / B_L \rceil + 1$ boundaries (one between each pair of adjacent blocks, plus the initial input and final output). Further optimization through sparse checkpointing is discussed in Section~\ref{subsec:storage_optimization}.

\subsection{Map-Reduce Hashing Scheme}
\label{subsec:hashing}

To bind recorded boundary states to tamper-evident commitments, we compute cryptographic hashes over boundary tensors during training. Any tampering with stored states will produce a hash mismatch during verification, ensuring that the recorded execution is cryptographically bound to its commitments.

However, computing cryptographic hashes sequentially over large tensors (e.g., millions of elements) creates a performance bottleneck that would significantly degrade training throughput. 

To address this, \sysname employs a map-reduce hash design optimized for accelerator (e.g., GPU) parallelism. The key idea is to partition each tensor into fixed-size chunks, compute hashes for all chunks in parallel on the accelerators, and then aggregate the chunk hashes into a single final hash.

For a tensor $X \in \mathbb{R}^{d_1 \times d_2 \times \cdots \times d_n}$ with $n_X = \prod_{i=1}^n d_i$ elements, the hashing procedure operates as follows:

\begin{enumerate}
    \item \textbf{Flatten and partition}: Flatten $X$ into a one-dimensional array and divide it into $m = \lceil n_X / C \rceil$ chunks of fixed size $C$.
    
    \item \textbf{Parallel hash}: Compute hashes for all chunks in parallel on accelerators, where each chunk is assigned to a separate thread:
    \begin{equation}
    h_k = H(\text{chunk}_k) \quad \text{for } k = 0, 1, \ldots, m-1,
    \end{equation}
    where $H$ is a cryptographic hash function such as BLAKE3 or SHA-256.
    
    \item \textbf{Aggregate}: Compute the final hash on the CPU by hashing the concatenation of all chunk hashes:
    \begin{equation}
    H(X) = H(h_0 \| h_1 \| \cdots \| h_{m-1}).
    \end{equation}
\end{enumerate}

The aggregate step ensures that the final hash depends on all chunks and their ordering, so any modification to $X$ produces a different hash value.

Applying this hashing scheme to the boundary states defined in Section~\ref{subsec:block_structure}, for each $\text{Block}_{i,j}$, we compute and record:
\begin{align}
\mathcal{R}_{i,j} = &\big\{ H(x_{i \cdot B_L, t}), H(y_{(i+1) \cdot B_L - 1, t}), \nonumber \\
&\quad H(\nabla x_{i \cdot B_L, t}), H(\nabla y_{(i+1) \cdot B_L - 1, t}) \big\}_{t \in \text{SB}_j} \nonumber \\
&\cup \big\{ H(\theta_{l,j \cdot B_S}), H(s_{l,j \cdot B_S}), \nonumber \\
&\quad H(\theta_{l,(j+1) \cdot B_S}), H(s_{l,(j+1) \cdot B_S}) \big\}_{l \in \text{LB}_i}.
\end{align}
These hashes form the tamper-evident audit trail that binds the recorded training execution to its commitments.

\subsection{Recomputation-Based Verification Protocol}
\label{subsec:verification}
With the block structure and cryptographic hashes established, we now describe how clients verify the integrity and correctness of training and inference. The verification protocol employs recomputation. Given model parameters and boundary states, the client re-executes the computation within the TEE and compares recomputed tensors against recorded hashes and values. Hash matching confirms data integrity, while numerical matching within floating-point tolerance confirms computational correctness. Algorithm~\ref{alg:verify} in Appendix~\ref{app:verify_alg} presents the verification protocol for training block cell $\text{Block}_{i,j}$.

Hash matching confirms bitwise integrity, where matching hashes indicate no tampering with recorded data. Numerical verification accounts for non-determinism in floating-point operations across hardware or software configurations. The relative $L_2$ error is compared against a threshold $\tau$ determined by the precision format (e.g., float32, bfloat16) and acceptable error bounds. This tolerance-based approach is essential for seamless integration. By accommodating legitimate floating-point variations, \sysname operates on standard infrastructure without requiring deterministic hardware, specialized frameworks, or modifications to existing ML pipelines. The effectiveness of verifying individual blocks may seem counterintuitive, especially considering that the hashes used for verification are computed by the service provider rather than the client. We provide a detailed explanation of why this approach is valid in Appendix~\ref{app:block_effectiveness}.

Importantly, block verification is inherently parallelizable, as each block cell depends only on its boundary states and can be verified independently. When verifying multiple blocks, clients can distribute the workload across multiple TEE instances, enabling scalable verification that grows linearly with available TEE resources rather than requiring sequential processing of the entire training trace.

\subsection{Inference Verification Protocol}
\label{subsec:inference_verification}

Inference verification applies to both fine-tuned models (after training) and base models (when clients rely on the provider solely for inference). \sysname extends the verification protocol to inference workloads to prevent model substitution or output fabrication.

Inference verification follows the same block structure and verification strategy as training, with two differences. First, only forward passes are recorded and recomputed, as inference involves no backpropagation or parameter updates. Second, each inference request is treated as a single step ($B_S = 1$). For each inference request with input $\mathbf{x}$, the provider records boundary activations $x_{i \cdot B_L}$ and $y_{(i+1) \cdot B_L - 1}$ for each layer block $\text{LB}_i$, along with their cryptographic hashes.

The client verifies an inference request by loading the selected block into TEE memory and recomputing the forward pass. The recomputed activations and their hashes are compared against the recorded values using the same hash matching and numerical verification procedures described in Section~\ref{subsec:verification}. Any attempt to substitute the model or fabricate outputs will produce mismatches in the intermediate or final prediction, enabling detection. This ensures end-to-end auditability, as clients can verify not only that training was executed correctly, but also that inference outputs are genuine predictions from the contracted model.

\subsection{Storage Optimization via Sparse Checkpointing}
\label{subsec:storage_optimization}

The design presented so far assumes that model parameters and optimizer states are saved at every step block boundary. While this provides direct verification, it may impose significant storage costs for long training runs.

To address this, \sysname supports sparse checkpointing strategies that trade storage for verification-time recomputation. Missing checkpoints or activations can be reconstructed via recomputation from nearby stored states. This enables flexible storage-computation trade-offs tailored to different deployment scenarios.

For training verification, we introduce the checkpoint interval $I_C$ to control the frequency of checkpointing along the step dimension. A checkpoint is saved at step block $\text{SB}_j$ if and only if $j \bmod I_C = 0$. Each checkpoint comprises:
\begin{equation}
\text{Checkpoint}_t = \{\theta_t, s_t\},
\end{equation}
where $t$ denotes the step at boundary $j \cdot B_S$. When $I_C > 1$, checkpoints are saved only at sparse intervals. 

To verify a step block $\text{SB}_j$ for which no checkpoint was saved, the system locates the nearest prior checkpoint and recomputes intervening steps to reconstruct the model state at the beginning of $\text{SB}_j$. This recomputation executes forward passes, backward passes, and parameter updates for each intervening step. For the TEE to verify the reconstructed state via hash matching, recomputation must produce bitwise-identical results, requiring deterministic computation and consistent hardware/software configurations. The reconstructed state is then provided to the TEE for verification.

Sparse checkpointing assumes deterministic computation, but certain layers may contain operations lacking deterministic implementations on some accelerators. \sysname addresses this through flexible layer block design, where non-deterministic layers can be isolated into separate layer blocks with $I_C = 1$, ensuring checkpoints are saved at every step boundary for these blocks. This selective strategy preserves verification correctness while enabling aggressive sparsity for deterministic layers.

With sparse checkpointing, the total storage cost for training becomes:
\begin{align}
\text{Storage}_{\text{total}} = &\frac{\lceil T / B_S \rceil}{I_C} \cdot (|\theta| + |s|) \nonumber \\
&+ (\lceil L / B_L \rceil + 1) \cdot T \cdot (|x| + |\nabla x|),
\end{align}
where $|\theta|$ and $|s|$ denote the sizes of parameters and optimizer states, and $|x|$ and $|\nabla x|$ denote the sizes of activations and gradients. The first term represents checkpoint storage (reduced by factor $I_C$), and the second term represents activation and gradient storage at layer block boundaries.

For inference verification, where model parameters remain fixed, an alternative optimization applies an activation interval $I_A$ along the layer dimension, where boundary states for layer block $\text{LB}_i$ are saved only if $i \bmod I_A = 0$. Missing activations are reconstructed by recomputing forward passes from adjacent recorded blocks. 

\subsection{Sampling-Based Spot-Checking and Strategies}
\label{subsec:sampling_based_spot_check}

While clients can verify all blocks for maximum assurance, \sysname supports selective verification through sampling to reduce overhead while maintaining security guarantees. Sampling-based spot-checking has been successfully applied to verify the integrity of data stored in the cloud~\cite{Ateniese2007datacloudstorage, Shacham2013datacloudstorage}. The key question is how effective probabilistic spot-checking is for verifying computational processes.

Let $N = \lceil L / B_L \rceil \times \lceil T / B_S \rceil$ denote the total number of block cells in the training or inference trace. If an adversary compromises $k$ blocks and the client samples a fraction $\pi$ of blocks for verification, the probability that the adversary evades detection is
\begin{equation}
P_{\text{evade}} = \frac{\binom{N - k}{\lfloor \pi N \rfloor}}{\binom{N}{\lfloor \pi N \rfloor}} = \prod_{i=0}^{\lfloor \pi N \rfloor - 1} \frac{N - k - i}{N - i}.
\end{equation}
The detection probability $P_{\text{detect}} = 1 - P_{\text{evade}}$ can be approximated as
\begin{equation}
P_{\text{detect}} \approx 1 - (1 - \pi)^{k} \approx 1 - e^{-\pi k}.
\end{equation}
Even modest verification efforts achieve substantial detection rates. For instance, if an adversary compromises $10\%$ of blocks ($k = 100$) out of $N = 1000$ and the client samples only $\pi = 1\%$ of blocks, the detection probability reaches $P_{\text{detect}} \approx 65\%$.

In addition, the single-round probabilities still understate the true deterrent power of spot-checking. The critical insight is that the adversary must execute the attack without knowing which blocks or how many will later be verified. This uncertainty forces the adversary to either avoid tampering entirely or risk detection, as any tampered block may be selected for verification. A single detection permanently destroys trust, which creates a powerful deterrent effect.

Unlike static data, where all segments are homogeneous, computational processes exhibit distinct characteristics across different stages and components. Beyond uniform random sampling, clients can thus adapt strategies to their specific threat priorities. Most importantly, the attacker cannot know the client's priorities and verification strategy at the time of the attack. For backdoor and poisoning attacks, the client need only verify the input block at each training step. For training degradation attacks such as epoch reduction or direct use of the base model, the existence of commitments already proves that training occurred. Randomly verifying a small number of blocks is sufficient to detect such attacks. For whole-model substitution attacks, selecting one block across the layer dimension at each training step is enough. These targeted blocks constitute only a small fraction of all blocks, yet provide near-certain detection for most such attacks. This strategic flexibility enables clients to achieve high detection rates with minimal verification cost, tailored to their specific threat models.

\subsection{Verification Timing}

\label{subsec:attacker_knowledge}
Security relies on keeping the client's future sampling choices hidden when the server submits its commitments. The client may choose any sampling strategy, such as uniform or biased sampling, and instantiate it with local randomness. The key distinction is whether the server has already submitted its commitments before it can observe the client's sampling choices or any side channel derived from verification activity.

\textbf{Offline verification.} All commitments are submitted before the client selects any block to audit. The server observes neither the client's sampling choice nor any side channel derived from verification activity, so it has no basis to predict which blocks will be checked. 

\textbf{Online verification.} The client begins auditing before training finishes, so early TEE invocation patterns can leak partial information about how often verification occurs and how blocks are sampled. The server may use this information to adjust the intensity of corruptions, and, when stable positional preferences are present, also shift corruptions toward blocks that appear less likely to be checked. This can weaken the security guarantee compared with offline verification. To reduce this risk, the client can switch sampling strategies frequently enough that past observations become stale, and use uniform sampling when it wants to avoid exposing any positional preference.

\textbf{Adaptive attacker.} The key limitation for the attacker, regardless of verification timing, is that for any block, the decision of whether it will be audited is made by the client's future sampling choice and is therefore unknown at the time the commitment must be submitted. However, online verification can additionally reveal partial information about the sampling strategy through TEE invocation patterns, giving the attacker a probabilistic advantage that offline verification removes entirely. Offline verification is therefore preferable whenever feasible. Under random sampling, the attacker's best strategy is to concentrate corruptions into as few blocks as possible, minimizing $k$ and reducing the probability that any sampled block has been tampered with. But any corrupted block remains fully detectable once verified, so detection always has positive probability.
\section{Evaluation}
\label{sec:evaluation}

We summarize the experimental setup in Appendix~\ref{app:exp_setup} and report the main evaluation results below.

\begin{table*}[ht]
\centering
\caption{Comprehensive system overhead comparison per 2048 samples. Prep time is the time to prepare blocks for verification. Verify time is per block cell in TEE.}
\label{tab:comprehensive_overhead}
\begin{tabular}{llcccccc}
\hline
\textbf{Model} & \textbf{Method} & \textbf{Block Size} & \textbf{Train Time} & \textbf{Overhead} & \textbf{Storage} & \textbf{Prep Time} & \textbf{Verify Time} \\
& & & \textbf{(s)} & & \textbf{(GB)} & \textbf{(s)} & \textbf{(s)}\\
\hline
\multirow{5}{*}{Qwen3-14B} 
& Baseline & - & 114.9 & - & - & - & - \\
& Step TEE & - & 188.3 & 64\% & 552.9 & 144.7 & N/A$^1$ % / TODO
\\
& \sysname & $B_L=2, B_S=4$ & 146.6 & 28\% & 303.8 & 13.0 & 29.8 % / TODO
\\
& \sysname & $B_L=4, B_S=8$ & 134.9 & 17\% & 206.4 & 28.0 & 77.5 % / TODO
\\
& \sysname & $B_L=4, B_S=16$ & 128.1 & 11\% & 178.9 & 39.8 & 122.1 % / TODO
\\
\hline
\multirow{5}{*}{Llama-3.1-8B} 
& Baseline & - & 62.8 & - & - & - & - \\
& Step TEE & - & 114.8 & 83\% & 333.3 & 78.0 & N/A$^1$ % / 178.5
\\
& \sysname & $B_L=2, B_S=4$ & 85.7 & 36\% & 201.0 & 7.3 & 21.0 % / 13.5
\\
& \sysname & $B_L=4, B_S=8$ & 76.1 & 21\% & 146.2 & 15.1 & 48.2 % / 32.6
\\
& \sysname & $B_L=4, B_S=16$ & 74.0 & 18\% & 131.3 & 24.5 & 76.2 % / 44.5
\\
\hline
\multirow{5}{*}{DINOv2-Giant} 
& Baseline & - & 24.1 & - & - & - & - \\
& Step TEE$^2$ & - & 87.6 & 263\% & 272.3 & - & N/A$^1$ % / 28.2
\\
& \sysname & $B_L=4, B_S=4$ & 44.1 & 83\% & 35.2 & 4.5 & 9.4 % / 5.6
\\
& \sysname & $B_L=8, B_S=8$ & 34.8 & 44\% & 19.1 & 10.4 & 25.6 % / 14.6
\\
& \sysname & $B_L=16, B_S=16$ & 30.1 & 25\% & 9.8 & 23.2 & 74.1 % / 48.0
\\
\hline
\multirow{6}{*}{ViT-Large} 
& Baseline & - & 6.8 & - & - & - & - \\
& Full TEE & - & 210.7 & 3000\% & - & - & - \\
& Step TEE & - & 18.2 & 168\% & 9.6 & 3.1 & 19.7 % / 12.7
\\
& \sysname & $B_L=8, B_S=4$ & 11.6 & 71\% & 6.5 & 3.4 & 8.7 % / 6.4
\\
& \sysname & $B_L=16, B_S=8$ & 8.9 & 31\% & 2.9 & 4.6 & 15.3 % / 9.2
\\
& \sysname & $B_L=16, B_S=16$ & 8.6 & 26\% & 2.3 & 7.6 & 21.7 % / 14.2
\\
\hline
\multicolumn{8}{l}{\small $^1$ Step TEE verification cannot be performed in TEE due to excessive memory consumption. CPU verification times are} \\
\multicolumn{8}{l}{\small \hspace{0.5em} 318.8s (Qwen3-14B), 178.5s (Llama-3.1-8B), and 28.2s (DINOv2-Giant). TEE time is estimated at 1.5--2$\times$ CPU time.} \\
\multicolumn{8}{l}{\small $^2$ DINOv2-Giant contains modules with non-deterministic operations. \sysname isolates these layers into separate blocks with} \\
\multicolumn{8}{l}{\small \hspace{0.5em} $I_C=1$, while Step TEE cannot handle layer-wise separation and thus cannot use checkpointing.} \\
\end{tabular}
\end{table*}

\subsection{System Overhead Analysis}
\label{subsec:overhead}

% \subsubsection{Training Overhead}
% \label{subsec:training_overhead}

Table~\ref{tab:comprehensive_overhead} presents overhead comparisons across different models, methods, and block configurations. We report training time, storage cost per 2048 samples, block preparation time for verification, and TEE verification time per block cell. Block preparation time refers to the time required to reconstruct model states for a target block from the nearest prior checkpoint via recomputation (Section~\ref{subsec:storage_optimization}), measured at the worst-case recomputation distance for sparse checkpointing with interval $I_C$.

Larger models incur lower relative training overhead, as hash computation cost becomes proportionally smaller with increasing model size. Full TEE training remains impractical even for small models due to substantial overhead, and memory constraints prevent larger LLMs from being loaded. In contrast, \sysname achieves practical overhead through block partitioning.

Block size configuration directly affects training overhead and storage cost, as larger blocks reduce both metrics by decreasing the number of hash computations and boundary recordings. Compared to Step TEE verification, \sysname reduces storage costs by over half across all models. Importantly, block verification can be performed in parallel without occupying GPU resources. %The storage cost can be further managed, which we discuss in Section~\ref{subsec:storage_management}.

Figure~\ref{fig:overhead_breakdown} in Appendix~\ref{app:overhead_breakdown} presents training overhead composition across different models using the second configuration from Table~\ref{tab:comprehensive_overhead}. Hash computation typically accounts for the majority of time overhead, as it requires GPU resources and stalls training until completion. Storage overhead composition varies by model architecture, where some models are parameter-dominant while others have larger activation and gradient storage.

Inference overhead results are reported in Appendix~\ref{app:inference_overhead}. Results for parameter-efficient fine-tuning with LoRA are reported in Appendix~\ref{subsec:lora_finetuning}, and a comparison between SGD and AdamW optimizers is provided in Appendix~\ref{subsec:optimizer_comparison}.

\begin{figure*}[t]
\centering
\includegraphics[width=0.73\linewidth]{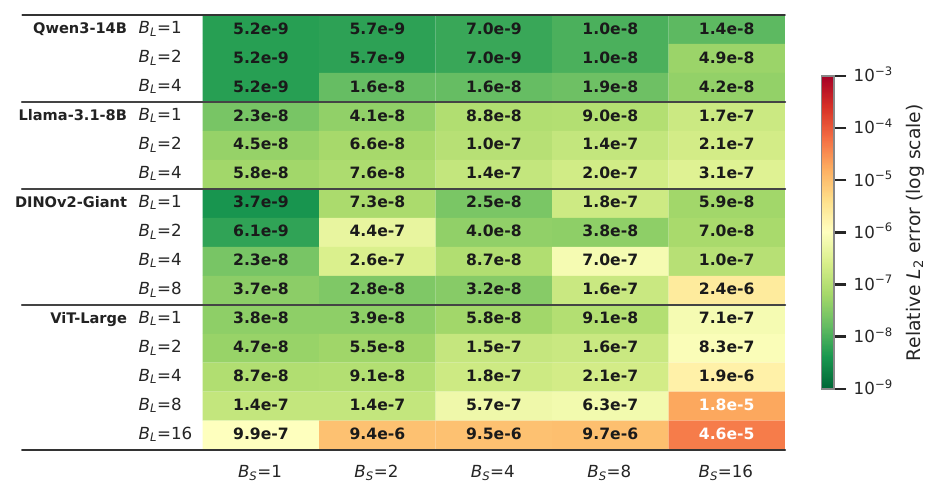}
\caption{
Training verification accuracy: parameter relative $L_2$ error across layer block sizes and step block sizes.
}
\label{fig:verification_accuracy}
\end{figure*}

\subsection{Verification Correctness}
\label{subsec:verification_correctness}

We examine whether recomputation-based verification produces numerically accurate results for both inference and training. 

\textbf{Training verification.} We measure the relative $L_2$ error of recomputed parameters against expected values. Figure~\ref{fig:verification_accuracy} reports the maximum layer-level parameter error for different layer block sizes ($B_L$) and step block sizes ($B_S$) using bfloat16 precision. These errors arise from floating-point computation differences between training and verification devices.

Numerical errors remain small across all configurations, well within acceptable tolerance for verification. The results reveal a clear error propagation trend, where larger $B_L$ and $B_S$ lead to higher errors, as longer recomputation sequences accumulate more floating-point discrepancies. This reflects the fundamental trade-off between training overhead and verification accuracy. Larger blocks reduce training overhead by minimizing hash computation frequency but require longer recomputation chains that amplify numerical errors. 

\begin{figure}[ht]
\centering
\includegraphics[width=0.95\linewidth]{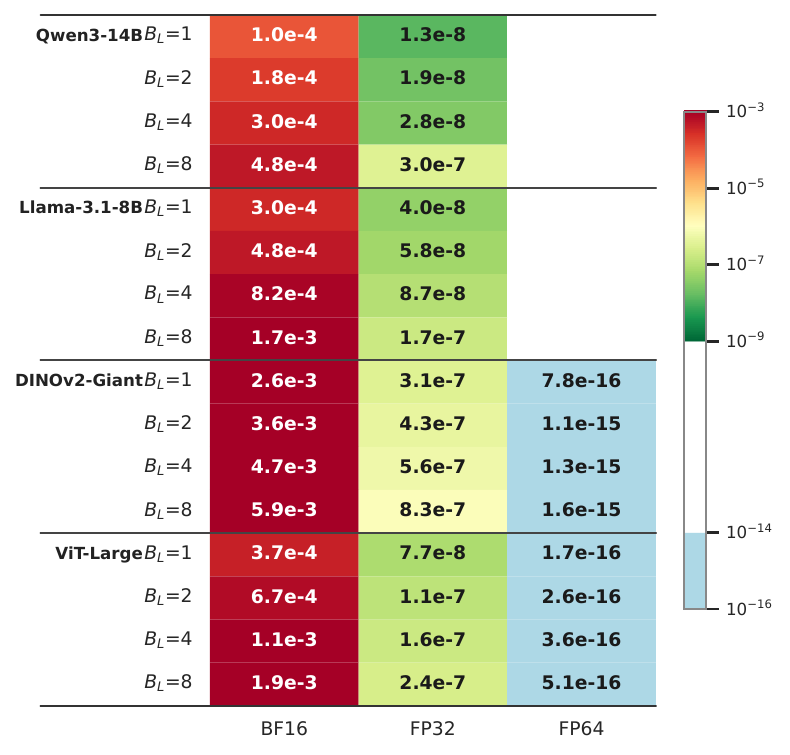}
\caption{
Inference verification accuracy: activation relative $L_2$ error across layer block sizes and precision levels.
}
\label{fig:inference_accuracy}
\end{figure}

\textbf{Inference verification.} Figure~\ref{fig:inference_accuracy} shows the relative $L_2$ error between recomputed and recorded activations for inference verification across different models and layer block sizes, comparing bfloat16, float32, and float64 precision.

Similar to training verification, a larger $B_L$ leads to higher errors due to longer recomputation sequences. However, bfloat16 precision produces errors too large for meaningful verification guarantees, leaving a tolerance gap that attackers can exploit, as we demonstrate in Section~\ref{subsec:adversarial_activation_attacks}. In contrast, float32 precision reduces these errors by 3-4 orders of magnitude, providing reliable verification guarantees that effectively detect malicious manipulations.

\subsection{Sampling Rate and Detection Probability}
\label{subsec:sampling_detection}

\begin{figure}[t]
\centering
\includegraphics[width=0.95\linewidth]{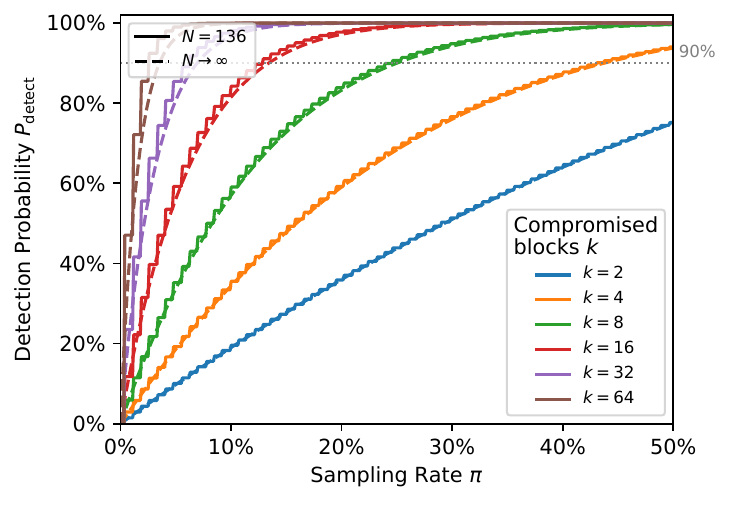}
\caption{Detection probability $P_{\text{detect}}$ as a function of sampling rate $\pi$ for different numbers of compromised blocks $k$.}
\label{fig:sampling_detection}
\end{figure}

As formalized in Section~\ref{subsec:sampling_based_spot_check}, when the client samples a fraction $\pi$ of blocks for verification, the detection probability against an adversary compromising $k$ blocks is $P_{\text{detect}} \approx 1-(1-\pi)^k$. Thus the effectiveness of a fixed sampling rate is determined by the smallest number of block cells that an attack must touch. We assume the adversary knows the system design and block division, so a rational adversary concentrates modifications within as few blocks as possible. Figure~\ref{fig:sampling_detection} illustrates how $P_{\text{detect}}$ grows with $\pi$ for representative values of $k$.

We therefore derive a conservative $k$ for each attack by counting the minimum block cells that must differ from an honest execution. Let $N$ denote the total number of block cells, $\alpha$ the attack strength, $L$ the total number of layers, $B_L$ and $B_S$ the layer and step block sizes, $M$ the number of training samples, and $S_b$ the batch size.

\textbf{Under-training attack.} The adversary skips a fraction $\alpha$ of training steps. Skipping steps changes the trace across all block cells associated with those steps, so the compromised block count scales with the full trace as $k_{\text{skip}} = \alpha \cdot N$.

\textbf{Data poisoning attack.} The adversary replaces a fraction $\alpha$ of training samples with malicious inputs, contaminating $\lceil \alpha M / S_b \rceil$ steps. Since the tampering occurs at the data input, the compromised cells are the input cells for those steps, giving $k_{\text{poison}} = \lceil \alpha M / (S_b \cdot B_S) \rceil$. A client targeting this attack should therefore prioritize input blocks rather than sample uniformly over the full trace.

For example, the original P-Trojan experiment~\cite{cui2025persistentbackdoorattackscontinual} uses 1401 poisoned training samples, giving a concrete scale for this calculation. With $B_S = 4$, these samples span at least 22 step-blocks. Uniformly verifying 10\% of the relevant input blocks achieves a 90\% detection rate in this case.

\textbf{Model substitution attack.} The adversary replaces a fraction $\alpha$ of layers at a training step. Since each block contains $B_L$ consecutive layers, replacing $L \cdot \alpha$ layers compromises at least $k_{\text{subst}} = \lceil L \cdot \alpha / B_L \rceil$ blocks.

\textbf{Parameter tampering attack.} The adversary modifies a fraction $\alpha$ of model parameters without altering the architecture. The compromised block count follows the same formula, $k_{\text{param}} = k_{\text{subst}}$. 

These $k$ values give the lowest detection probability for each attack under our sampling model. Table~\ref{tab:sampling_detection} reports the resulting detection probabilities for Llama-3.1-8B with 2048 training samples.

\begin{table*}[t]
\centering
\caption{Detection probability under different attack types, block configurations, and sampling strategies (Llama-3.1-8B, 2048 training samples).}
\label{tab:sampling_detection}
\begin{tabular}{llccccc}
\hline
\textbf{Attack Type} & \textbf{Attack Strength} & $B_L$ & $B_S$ & \textbf{Strategy} & $\pi$ & $P_{\text{detect}}$ \\
\hline
\multirow{2}{*}{Under-training}
 & Skip 50\% steps & \multirow{2}{*}{4} & \multirow{2}{*}{8} & \multirow{2}{*}{Uniform} & \multirow{2}{*}{5\%} & 99.3\% \\
 & Skip 25\% steps & & & & & 87.3\% \\
\hline
\multirow{4}{*}{Data poisoning}
 & \multirow{4}{*}{10\% poison rate} & 4 & 8 & Uniform    & 20\% & 35.9\% \\
 & & 4 & 8 & Input-only & 10\% & 97.8\% \\
 & & 4 & 4 & Uniform    & 20\% & 59.0\% \\
 & & 4 & 4 & Input-only & 10\% & 99.9\% \\
\hline
\multirow{4}{*}{Model substitution}
 & \multirow{2}{*}{Replace 100\% layers} & 4 & 8 & \multirow{4}{*}{Uniform} & \multirow{4}{*}{20\%} & 87.3\% \\
 & & 2 & 8 & & & 98.0\% \\
 & \multirow{2}{*}{Replace 75\% layers}  & 4 & 8 & & & 79.6\% \\
 & & 2 & 8 & & & 94.9\% \\
\hline
\multirow{4}{*}{Parameter tampering}
 & \multirow{2}{*}{Modify 20\% params} & 4 & 8 & \multirow{4}{*}{Uniform} & \multirow{2}{*}{30\%} & 51.4\% \\
 & & 2 & 8 & & & 76.3\% \\
 & \multirow{2}{*}{Modify 10\% params} & 4 & 8 & & \multirow{2}{*}{50\%} & 50.0\% \\
 & & 2 & 8 & & & 74.9\% \\
\hline
\end{tabular}
\end{table*}

Under-training is the easiest attack to detect because skipped steps produce a large $k$ across the full block trace. Data poisoning has a smaller full-trace footprint, but its compromised cells are localized to inputs, so input-only sampling gives much higher detection probability than uniform sampling at lower cost. Model substitution and parameter tampering are harder to detect when the modification is confined to only one step, which keeps $k$ small.

\subsubsection{Practical Guidance}
\label{subsubsec:practical_guidance}
Block configuration directly shapes detection probability. A finer $B_L$ increases the number of blocks layer-wise and thus $k$ for layer-level attacks, while a finer $B_S$ increases $k$ for step-level attacks such as data poisoning. However, Section~\ref{subsec:overhead} also shows that finer blocks increase training overhead. Clients should therefore consider both their target threat model and the overhead budget when negotiating block configuration with the service provider, treating the choice of $B_L$ and $B_S$ as a security parameter alongside the sampling rate $\pi$.

When selecting $\pi$, clients should reason about both the attack type they wish to defend against and the plausible attack strength $\alpha$. For under-training, even a modest sampling rate suffices. For model substitution, clients can estimate the minimum $\alpha$ that would make the attack worthwhile to the provider. Replacing only a small fraction of layers with a weaker model likely provides insufficient benefit. Clients can therefore derive a lower bound on $k_{\text{subst}}$ from that threshold and choose $\pi$ accordingly. When the threat model assumes an actively malicious provider, such as one carrying out data poisoning or parameter tampering, a conservative client should adopt a higher $\pi$. For attacks whose compromised blocks are predictable, such as data poisoning, where tampering is confined to input-associated cells, the client can apply a biased sampling strategy that concentrates verification on those blocks, substantially reducing the required $\pi$ while maintaining high detection probability. In this case, as discussed in Section~\ref{subsec:attacker_knowledge}, the client should delay verification until training has fully completed. If the client is extremely risk-averse, it may opt to verify all blocks to achieve deterministic detection.

\subsubsection{False Positives}
\label{subsubsec:false_positive}
For most attack types, the false positive rate is zero by construction. Under-training requires fabricating hashes for blocks that were never computed, which hash matching rejects outright. Model substitution produces dimension mismatches in saved parameters, detected structurally before any numerical comparison. Data poisoning requires altering the training dataset after the client has already hashed it, producing an immediate hash mismatch.

Parameter tampering is the one case where the client must judge whether the TEE recomputation result agrees with the provider's recorded output within floating-point tolerance. Here a false positive would mean flagging a legitimate numerical discrepancy as an attack. Appendix~\ref{app:backdoor_defense} evaluates BadEdit~\cite{li2025backdoorllm, li2024badedit}, which uses triggers as backdoors, and Concept-ROT~\cite{grimes2025conceptrot}, which uses concepts as backdoors, showing that verifying the compromised blocks reveals large parameter deviations. Together with the empirical lower bound analysis in Section~\ref{subsec:tolerance_exploitation}, these results show that the output deviation introduced by meaningful parameter tampering exceeds normal floating-point recomputation error by several orders of magnitude. We therefore consider it unlikely that ordinary numerical precision differences would be misidentified as a malicious attack.

\subsection{Security Against Numerical Tolerance Exploitation}
\label{subsec:tolerance_exploitation}

Both inference and training verification rely on numerical error thresholds to accommodate legitimate floating-point differences across devices. A security concern is whether adversaries could exploit these tolerances to inject malicious perturbations that remain below detection thresholds yet compromise model behavior. We evaluate this threat through two attack vectors, adversarial examples on activations (inference-level) and parameter poisoning (training-level).

\subsubsection{Adversarial Example Attacks on Activations}
\label{subsec:adversarial_activation_attacks}

As established in Section~\ref{subsec:inference_verification}, \sysname prevents providers from fabricating outputs by verifying recorded activations. However, could adversaries inject adversarial perturbations into intermediate activations that evade detection?

\begin{figure}[ht]
\centering
\includegraphics[width=0.95\linewidth]{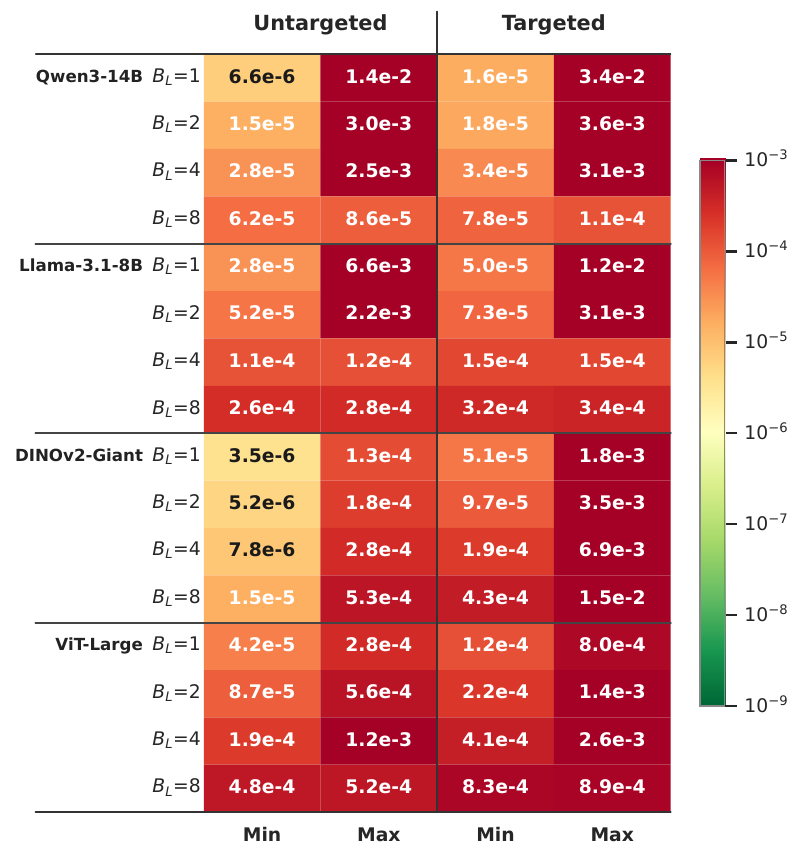}
\caption{
Adversarial example attacks: activation perturbation relative $L_2$ norm across models and layer block sizes.
}
\label{fig:adversarial_attack_stats}
\end{figure}

We formulate the untargeted and targeted activation attacks in Appendix~\ref{app:activation_perturbation_attacks}. We evaluate the attack with varying layer block sizes to understand how block granularity affects the perturbation budget required for successful attacks. Figure~\ref{fig:adversarial_attack_stats} summarizes the maximum and minimum $L_2$ perturbation magnitudes across all blocks for both untargeted and targeted attacks.

Maximum perturbations occur predominantly at early layers, while minimum perturbations concentrate at later layers, as later layers require smaller perturbations due to shorter propagation paths to the output. As $B_L$ increases, maximum perturbations decrease because recorded activations shift to later layers, while minimum perturbations increase due to fewer boundaries to exploit.

Comparing these perturbation magnitudes against numerical errors in Figure~\ref{fig:inference_accuracy} reveals a critical vulnerability when using bfloat16 precision. In many configurations, adversarial perturbations fall within or near bfloat16's recomputation error range, making it nearly impossible to establish an effective threshold that distinguishes malicious perturbations from legitimate numerical errors without triggering excessive false positives. This enables attacks that evade detection by exploiting the inherent numerical tolerance.

We note, however, that even when the precision setting leaves room for such perturbations, mounting this attack at inference time remains difficult in practice. Crafting an adversarial perturbation requires backpropagation through the model and solving an optimization problem, whereas inference requests issued by clients are typically expected to be served in real time. The provider therefore rarely has a sufficient time window to solve the optimization and still return the result within the expected latency.

In contrast to bfloat16, float32 keeps numerical errors orders of magnitude smaller than even the minimum adversarial perturbations across all configurations, providing a clear security margin for detection. Going further, float64 reduces numerical errors to negligible levels, offering more separation than verification needs. Since the latency argument above is a practical mitigation rather than a security guarantee, float32 should be regarded as the minimum precision for inference verification. Importantly, both precision levels are supported by standard GPU hardware, preserving \sysname's seamless integration with existing ML infrastructure.

\subsubsection{Parameter Poisoning Attacks on Model Weights}

Training verification records parameter hashes at block boundaries, subject to numerical recomputation errors. Could adversaries exploit these tolerances by injecting small parameter perturbations to compromise model behavior?

\begin{figure}[ht]
\centering
\includegraphics[width=0.95\linewidth]{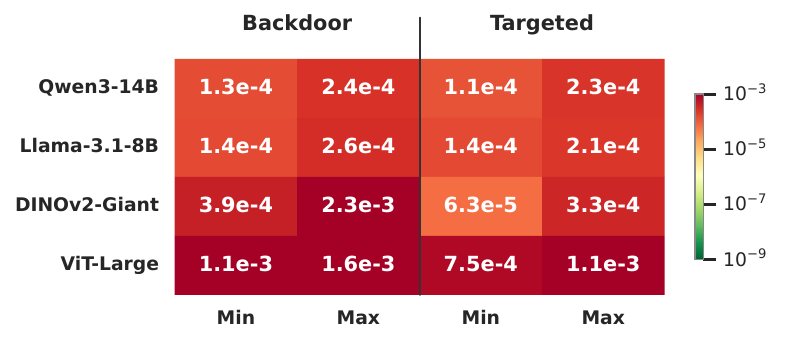}
\caption{
Parameter poisoning attacks: parameter perturbation relative $L_2$ norm across models.}
\label{fig:parameter_poisoning_stats}
\end{figure}

We formulate the backdoor and targeted parameter poisoning attacks in Appendix~\ref{app:parameter_poisoning_attacks}. We measure the minimum and maximum relative $L_2$ parameter perturbations required for successful attacks using bfloat16 precision. Figure~\ref{fig:parameter_poisoning_stats} summarizes the perturbation magnitudes for both backdoor and targeted attacks across different models.

Successful parameter poisoning requires perturbations orders of magnitude larger than numerical recomputation errors shown in Figure~\ref{fig:verification_accuracy}. Combined with activation-level results, this confirms that with proper precision choices, \sysname's numerical tolerances cannot be exploited to inject stealthy attacks at either inference or training level. The substantial separation between numerical errors and attack perturbations enables effective detection of meaningful tampering while accommodating legitimate floating-point variations across ML accelerators and TEE hardware.

% \section{Discussion}
% \label{sec:disscussion}

% \subsection{Storage Management}
% \label{subsec:storage_management}

% \sysname's verification storage is temporary: recorded states exist solely for audit and can be released immediately after verification completes. To further reduce storage burden, the provider can optionally adopt an incremental commit-and-prune strategy: periodically transmit hash records $\{\mathcal{R}_{i,j}\}$ for completed training stages to the client, who specifies target blocks for verification. The provider then deletes all non-target states before TEE provisioning, retaining only data for requested blocks. This optional strategy enables proactive storage reclamation during training rather than requiring full retention until verification completes.

% For extreme storage constraints, \sysname supports a zero-storage strategy ($I_C=\infty,I_A=\infty$) where only cryptographic hashes are recorded and transmitted during training, and the size of these commitments is negligible. Upon receiving the client's verification request, the provider reruns the entire training from the initial checkpoint, recording states only for the requested blocks. This trades 100\% additional training overhead (one full retraining pass) to eliminate temporary storage during initial training. Providers can adapt the system across a spectrum: from dense recording with minimal verification overhead to zero-storage with doubled training cost, matching their specific resource constraints, which demonstrates \sysname's flexibility in balancing storage and computation costs.

\section{Limitations}
Our work relies on TEEs as the root of trust for verification. Like all TEE-based approaches, if the TEE itself is compromised, the security guarantees of our framework will be undermined. In addition, when clients choose to verify only a subset of blocks, the framework provides probabilistic detection guarantees rather than deterministic assurance, so malicious behavior may still have a chance of evading detection.
\section{Conclusion}
\label{sec:conclusion}

Cloud-based fine-tuning and inference create a fundamental trust gap. Clients cannot verify whether providers execute computations as contracted, and existing verification approaches fail to scale to modern LLMs due to either prohibitive overhead (ZKP-based methods) or full model encapsulation requirements (TEE-based methods). \sysname addresses this through block decomposition with compositional verification. By recording commitments over boundary states and enabling selective TEE-based recomputation, \sysname provides strong auditability while keeping primary workloads on standard accelerators with low additional overhead.

Our evaluation demonstrates that the two-dimensional block structure offers flexibility to balance overhead, storage, and verification granularity, while map-reduce style hashing minimizes commitment costs. The sampling-based verification strategy provides probabilistic detection guarantees that compound over repeated audits. By decoupling execution from verification, \sysname transforms opaque AI services into accountable platforms where trust is established through verifiable evidence rather than provider claims.

\section*{Acknowledgments}

This work was supported in part by the Office of Naval Research under grants N00014-24-1-2730 and N00014-24-1-2663, and the National Science Foundation under grants 2433904, 2312447, 2247560, 2154929, 2235232, and 2238635, and the Army Research Office under grant W911NF-24-1-0155.

%Lou: N00014-24-1-2730 (FL), 2433904(NewSpectrum), 2312447 (NeTS), 2247560 (SaTC 3), 2154929 (SaTC 2), and 2235232 (CPS).
%Ning: NSF (CNS-2238635), ARO (W911NF-24-1-0155), and ONR (N000142412663)

\section*{Open Science}
All code and artifacts used for the experiments in this paper are available at \url{https://doi.org/10.5281/zenodo.20438270} and \url{https://github.com/hengvt/AFTUNE}, with detailed instructions for using the framework and reproducing the experiments.

\section*{Ethical Considerations}
We structure the ethical considerations by linking our stakeholder analysis to the impacts generated during the research process and publication of results.

\textbf{Stakeholder Analysis.} \sysname involves four primary stakeholder groups. (1) \textit{Cloud Service Providers}: Companies offering AI fine-tuning and inference services. (2) \textit{Clients}: Organizations outsourcing model fine-tuning and inference to cloud providers. (3) \textit{End Users}: Individuals who interact with AI systems built on cloud-fine-tuned models. (4) \textit{Research Community}: ML and security researchers who may build upon our framework.

\textbf{Positive Impacts.} (1) \textit{Enhancing Trust in Cloud AI (Impact on Clients \& End Users)}: \sysname provides clients with verifiable guarantees that contracted computations were executed faithfully, reducing the risk of silent service degradation or fraud. This indirectly benefits end users who rely on properly trained models. (2) \textit{Promoting Accountability (Impact on Providers \& Society)}: By enabling auditability, our framework incentivizes honest behavior among cloud providers, fostering a more trustworthy AI service ecosystem. (3) \textit{Reproducibility and Open Science (Impact on Research Community)}: We open-source our implementation to enable the community to assess, audit, and improve upon our methods.

\textbf{Negative Impacts.} We are not aware of any potential harms or negative societal impacts arising from this work. Our framework is purely defensive in nature, designed to detect misbehavior rather than enable it.

\textbf{Decision to Conduct and Publish.} We initiated this research to address a practical trust problem in cloud AI services. Our work focuses on defense rather than offense. We believe sharing these results benefits the broader ecosystem by enabling clients to hold providers accountable.

{\footnotesize \bibliographystyle{acm}
\bibliography{ref}}

\appendix

\section{Possible Attacks}
\label{app:attacks}

Below we outline several classes of provider misbehavior that the visibility gap makes possible. In all these cases, the attacks can be profitable for the provider and hard to catch with final accuracy or black-box testing alone, while leaving the client with no verifiable record of what actually happened during training or inference.

\subsection{Attacks in Training}

\subsubsection{Under-training}
An untrusted provider may try to cut training cost while still charging for the full job. The provider might silently reduce epochs, truncate batches, skip parts of the dataset, or drift from the agreed hyperparameter schedule, then bill as if the contracted training had been run in full. For example, a client who paid for 3 epochs over 100K samples might receive a model trained for 1 epoch on 50K samples, good enough to pass rough checks but weaker on held-out or long-tail data.

\subsubsection{Model substitution}
Instead of the contracted base model, the provider may use a weaker checkpoint from the start or switch to a cheaper model (e.g., a smaller or distilled one) partway through training. The client pays for fine-tuning from a strong base but has no way to verify that the delivered model was actually trained from that base.

\subsubsection{Backdoors and poisoning}
A compromised or malicious provider could inject backdoors or other latent behaviors during training. These pass normal quality checks but trigger on specific inputs (e.g., certain phrases or embedded triggers), so the model misbehaves only when those conditions are met. For example, a malicious employee at the provider could add a backdoor to the training set to steer the LLM's behavior, then remove that backdoor from the training set after training is done. As long as the trigger remains unknown, this behavior cannot be detected or verified by the client.

\subsection{Attacks in Inference}

\subsubsection{Serving the wrong model}
At serving time, the provider might run a cheaper model than the one the client believes is deployed. For instance, the provider may serve the base model or a distilled copy instead of the client's fine-tuned model, while claiming to serve the latter.

\subsubsection{Fabricated or manipulated responses}
The provider might skip running the model altogether for some requests, returning cached or templated responses for repeated or similar queries to save compute. A malicious or compromised provider may also directly manipulate returned results, for example, deliberately altering or injecting responses to steer the client's downstream decisions, spread misinformation, or exfiltrate or poison data.

\section{Algorithm to Verify One Training Block}
\label{app:verify_alg}

\begin{algorithm}
\caption{Verify Block Cell}
\label{alg:verify}
\begin{algorithmic}[1]
\Require Layer block index $i$, step block index $j$, recorded hash set $\mathcal{R}_{i,j}$, parameters $\{\theta_{l,j \cdot B_S}\}_{l \in \text{LB}_i}$, optimizer states $\{s_{l,j \cdot B_S}\}_{l \in \text{LB}_i}$, boundary states, and numerical tolerance $\tau$
\Ensure Verification result: \textsc{Pass} or \textsc{Fail}
\State \textbf{Step 1: Load model state in TEE}
\State Load parameters and optimizer states for layers in $\text{LB}_i$ at step $j \cdot B_S$ into TEE memory
\If{$H(\theta_{l, j \cdot B_S}) \notin \mathcal{R}_{i,j}$ or $H(s_{l, j \cdot B_S}) \notin \mathcal{R}_{i,j}$ for any $l \in \text{LB}_i$}
    \State \Return \textsc{Fail}
\EndIf
\State \textbf{Step 2: Verify target block cell}
\For{each step $t \in \text{SB}_j$}
    \State \textbf{Hash verification (integrity):}
    \State Load boundary states $x_{l_{\text{in}}, t}$, $y_{l_{\text{out}}, t}$, and $\nabla y_{l_{\text{out}}, t}$
    \If{$H(x_{l_{\text{in}}, t}) \notin \mathcal{R}_{i,j}$ or $H(y_{l_{\text{out}}, t}) \notin \mathcal{R}_{i,j}$}
        \State \Return \textsc{Fail}
    \EndIf
    \State \textbf{Forward pass:}
    \State $\hat{y}_{l_{\text{out}}, t} \gets \text{ForwardBlock}(\text{LB}_i, x_{l_{\text{in}}, t})$
    \State \textbf{Numerical verification:}
    \If{$\|\hat{y}_{l_{\text{out}}, t} - y_{l_{\text{out}}, t}\| / \|y_{l_{\text{out}}, t}\| > \tau$}
        \State \Return \textsc{Fail}
    \EndIf
    \State \textbf{Backward pass and gradient verification:}
    \State $\hat{\nabla} x_{l_{\text{in}}, t}, \hat{\nabla} \theta_t \gets \text{BackwardBlock}(\text{LB}_i, \nabla y_{l_{\text{out}}, t})$
    \State Verify hashes of loaded gradients and numerical correctness of recomputed gradients (similar to forward pass)
    \State \textbf{Parameter update and verification:}
    \State $\hat{\theta}_{t+1}, \hat{s}_{t+1} \gets \text{OptimizerStep}(\hat{\theta}_t, \hat{s}_t, \hat{\nabla}\theta_t)$
    \State Verify hashes and numerical correctness for updated parameters and optimizer states
\EndFor
\State \Return \textsc{Pass}
\end{algorithmic}
\end{algorithm}

\section{Verification Chain and Trust Anchors}
\label{app:block_effectiveness}

As described in Section~\ref{subsec:verification}, verification inside the TEE depends on hash matching, but the hashes themselves are computed by the service provider during training. If the provider is untrusted, how can its hashes serve as a basis for trust?

In traditional data storage auditing, hashes are computed by the client before uploading data, so the client can directly trust the hash values. In our framework, however, hashes are computed by the cloud provider during training. The correctness of these hashes is instead guaranteed through a verification chain formed by neighboring blocks.

\begin{figure}[h]
\centering
\includegraphics[width=1\linewidth]{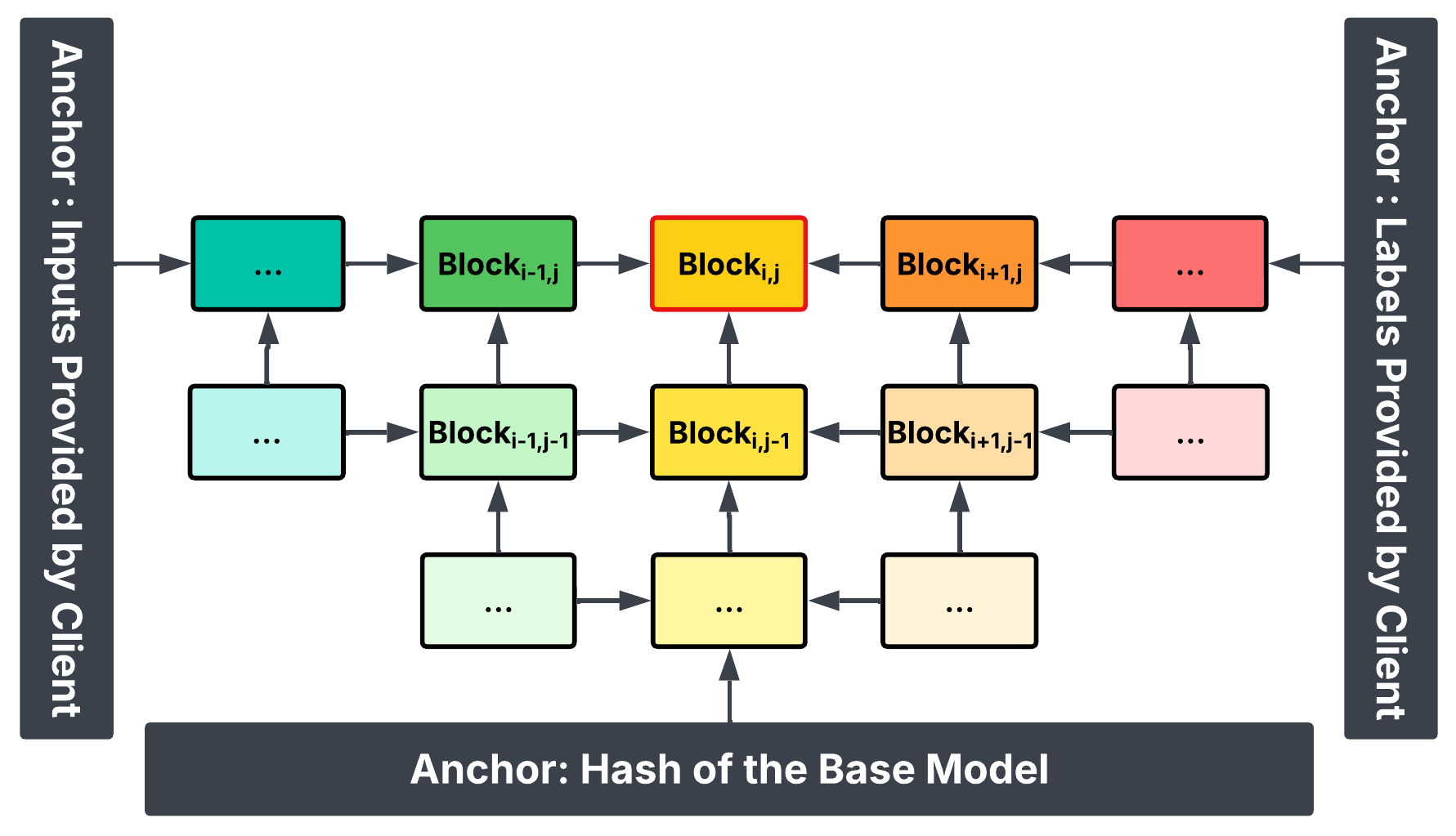}
\caption{
Trust propagation to $\text{Block}_{i,j}$. Correctness is derived from the anchor through neighboring blocks. Outgoing trust flows are omitted for clarity. Note that commitments of all blocks within each row must be submitted together, establishing mutual dependencies among them.}
\label{fig:trust_propagation}
\end{figure}

Specifically, for any $\text{Block}_{i,j}$, the hash values used for verification derive their correctness from adjacent blocks, where $\text{Block}_{i-1,j}$ guarantees the correctness of inputs, $\text{Block}_{i+1,j}$ that of gradients, and $\text{Block}_{i,j-1}$ that of parameters, following the model computation procedure. The correctness of these neighboring blocks can be similarly traced back to their own neighbors. Ultimately, all blocks form a verification chain, where each block serves to validate its adjacent blocks, and the entire chain is anchored to trusted roots, namely the initial model weights and the client-provided inputs and labels. Figure~\ref{fig:trust_propagation} illustrates this trust propagation.

We require the cloud provider to publicly release the hash commitment of the base model before any fine-tuning contract is established. This serves as the parameter anchor without revealing actual model weights. At this point, since no specific fine-tuning agreement exists, the provider cannot preemptively attack the base model for a particular client. The inputs and labels are provided by the client, who can compute their hashes independently. Thus, for any block, its hash correctness is ultimately linked through other blocks and anchored to these established trust roots. We reiterate that for each block, at the time when an attack could be performed, the attacker cannot know whether this block will later be selected for verification.

One might ask what happens if an attacker successfully compromises a block that happens not to be verified, as discussed in Section~\ref{subsec:sampling_based_spot_check}. If such a block escapes verification, the damage to that specific block has already occurred. However, since the hash commitments for this block were also submitted and remain immutable, they become new trust anchors for verifying neighboring blocks, and any new attack on other blocks would still be detected. Consequently, the verification of other blocks remains unaffected. Therefore, our framework provides the same security guarantees as traditional data storage auditing for the overall training process.

\section{Experimental Setup}
\label{app:exp_setup}

We evaluate \sysname across diverse model architectures to measure its overhead, storage efficiency, and verification accuracy.

\textbf{Models.} We evaluate \sysname on both language and vision models to demonstrate its generality. Llama-3.1-8B~\cite{llama3:meta24} and Qwen3-14B~\cite{qwen:alibaba} are used for language tasks, and DINOv2-Giant~\cite{dinov2} and ViT-Large~\cite{vitlarge} are used for vision tasks. These models span different sizes, layer structures, and computational patterns.

\textbf{Datasets.} For LLM fine-tuning, we use MentalChat16K~\cite{xu2025mentalchat16k}, a mental health conversational dataset. For vision models, we use ImageNet-1K~\cite{imagenet15russakovsky} for image classification tasks.

\textbf{Baselines.} We compare \sysname against three baselines. \textit{Baseline} is standard training without verification and represents the performance upper bound. \textit{Full TEE} executes the entire training process within a TEE, providing strong security but incurring significant overhead. \textit{Step TEE} executes verification in TEE but checks one complete training step at a time across all layers, lacking \sysname's flexible block-based granularity control. Due to memory constraints, Full TEE and Step TEE are only feasible for ViT-Large among our evaluated models. For other models, we execute on CPUs solely for overhead comparison.

\textbf{Configurations.} We vary block sizes $B_L$ and $B_S$, hash functions, optimizers, and checkpoint intervals $I_C$. The default configuration uses $B_L = 4$, $B_S = 8$, BLAKE3 hash function, SGD optimizer, batch size 16, bfloat16 precision, and $I_C = 8$ unless otherwise stated. For LLMs we use a learning rate of $1 \times 10^{-5}$, and for vision models we use $1 \times 10^{-4}$. For hash computation, we use chunk sizes 4096, 3072, 2048, and 1024 for Qwen, Llama, DINOv2, and ViT, respectively.

\textbf{Hardware and Environment.} Experiments run on NVIDIA RTX PRO 6000 GPUs with 96GB GPU memory and Intel Xeon Gold 5520+ CPUs for TEE verification with Intel SGX and TDX support. We implement the TEE verification component using Gramine~\cite{Gramine,Gramine-TDX}, a library OS that supports both Intel SGX and Intel TDX for hardware-backed attestation and isolation. The TEE memory is configured to 16GB, the maximum memory size supported by our platform.

\section{Training Overhead Composition}
\label{app:overhead_breakdown}

\begin{figure*}[ht]
\centering
\begin{minipage}{0.9\textwidth}
\centering
% Time overhead breakdown - top row with legend
\includegraphics[width=0.17\linewidth]{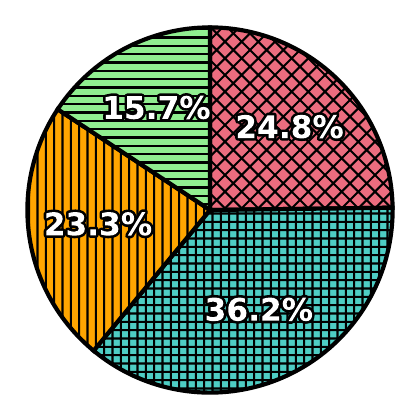}
\includegraphics[width=0.17\linewidth]{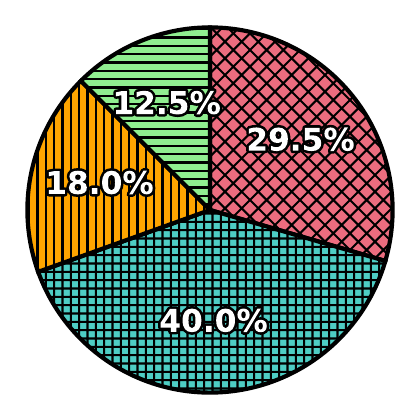}
\includegraphics[width=0.17\linewidth]{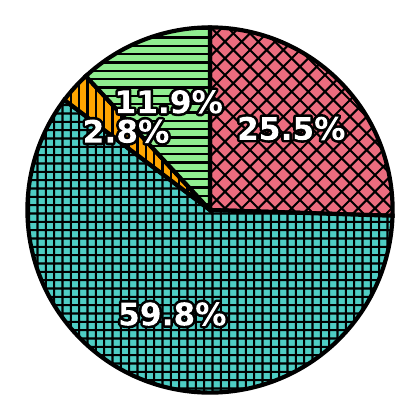}
\includegraphics[width=0.17\linewidth]{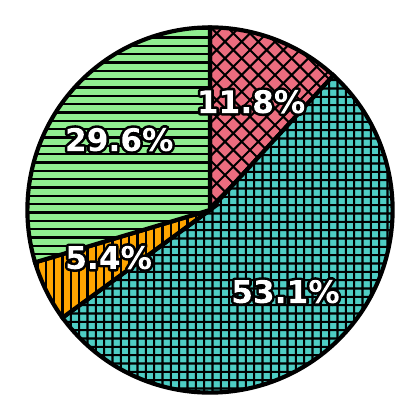}
\includegraphics[width=0.17\linewidth]{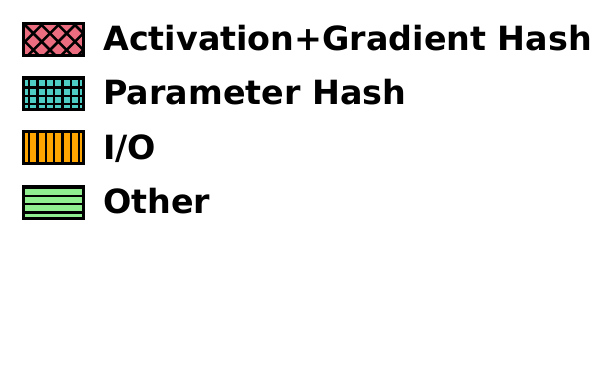}

% Storage breakdown - bottom row with legend
\includegraphics[width=0.17\linewidth]{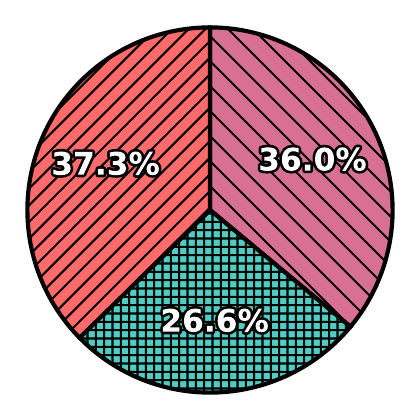}
\includegraphics[width=0.17\linewidth]{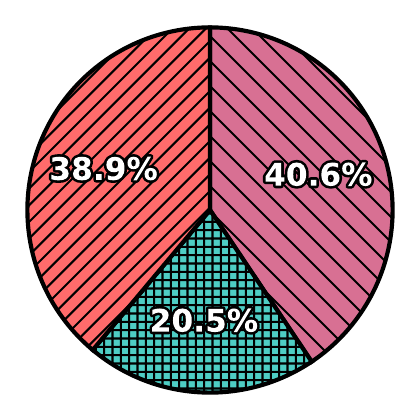}
\includegraphics[width=0.17\linewidth]{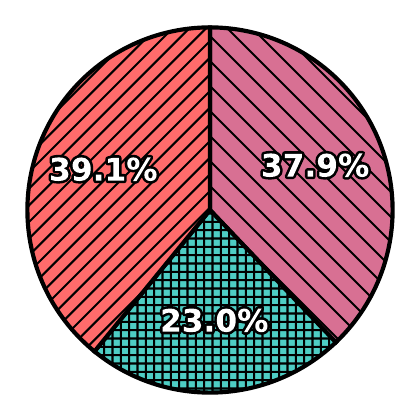}
\includegraphics[width=0.17\linewidth]{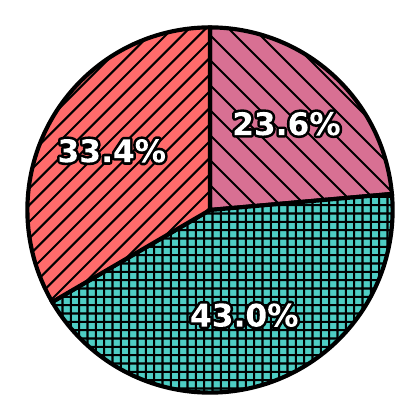}
\includegraphics[width=0.17\linewidth]{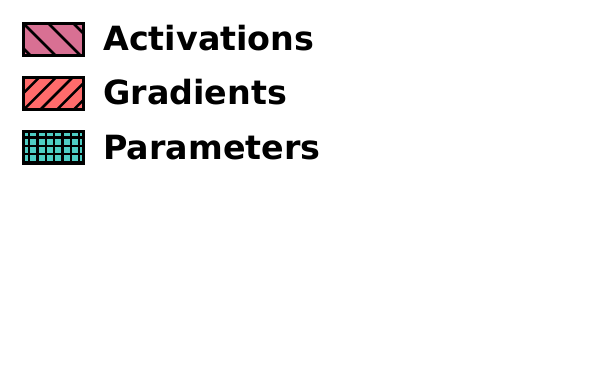}
\end{minipage}

\caption{Training overhead composition across different models with the second configuration from Table~\ref{tab:comprehensive_overhead}. \textbf{Top row}: Time overhead proportions for Qwen3-14B, Llama-3.1-8B, DINOv2-Giant, ViT-Large (left to right), followed by legend showing hash computation, I/O, and other \sysname operations. \textbf{Bottom row}: Storage proportions for the same models, followed by legend showing activations, gradients, and parameters.}
\label{fig:overhead_breakdown}
\end{figure*}

Figure~\ref{fig:overhead_breakdown} shows a breakdown of overhead across different types of operations. The ``Other'' overheads include system costs from memory management, synchronization waits, computation flow interruptions, and record maintenance.

\section{Inference Overhead}
\label{app:inference_overhead}
\label{subsec:inference_verification_exp}

We measure inference latency overhead, storage cost, and verification time per inference request. Table~\ref{tab:inference_overhead_bl} examines the impact of layer block size ($B_L$) with $I_A=4$.

\begin{table*}[ht]
    \centering
    \caption{Impact of layer block size on inference verification overhead with $I_A=4$. Rows with $B_L = $ ``-'' are baselines.}
    \label{tab:inference_overhead_bl}
    \begin{tabular}{cccccc}
    \hline
    \textbf{Model} & \textbf{$B_L$} & \textbf{Infer Time (ms)} & \textbf{Overhead} & \textbf{Storage (MB)} & \textbf{Verify Time (s)} \\
    \hline
    \multirow{4}{*}{Qwen3-14B}
    & - & 339 & - & - & - \\
    & 1 & 391 & 15\% & 3.16 & 17.5 \\
    & 2 & 386 & 14\% & 1.73 & 20.5 \\
    & 4 & 363 & 7\% & 0.88 & 27.0 \\
    \hline
    \multirow{4}{*}{Llama-3.1-8B}
    & - & 341 & - & - & - \\
    & 1 & 389 & 14\% & 3.75 & 6.3 \\
    & 2 & 372 & 9\% & 2.09 & 8.3 \\
    & 4 & 355 & 4\% & 0.84 & 12.5 \\
    \hline
    \multirow{4}{*}{DINOv2-Giant}
    & - & 267 & - & - & - \\
    & 1 & 331 & 24\% & 8.60 & 4.6 \\
    & 2 & 303 & 13\% & 4.82 & 4.9 \\
    & 4 & 296 & 11\% & 2.56 & 5.3 \\
    \hline
    \multirow{4}{*}{ViT-Large}
    & - & 212 & - & - & - \\
    & 1 & 266 & 25\% & 3.00 & 4.5 \\
    & 2 & 241 & 14\% & 1.84 & 4.7 \\
    & 4 & 241 & 14\% & 1.06 & 5.4 \\
    \hline
    \end{tabular}
    \end{table*}

Inference recording imposes modest latency overhead, around 10\% in optimal configurations. Storage requirements per inference request are highly practical, typically 1-5 MB depending on model size and block configuration. Verification time is also acceptable, taking 5-30 seconds depending on model complexity, allowing clients to audit inference outputs efficiently.

\section{Defense Against Backdoor Attacks}
\label{app:backdoor_defense}

To demonstrate the effectiveness of \sysname against backdoor attacks, we evaluate the two parameter-level backdoor attacks discussed in Section~\ref{subsubsec:false_positive} on Llama-3.1-8B with $B_L=2$ and $B_S=8$. For Concept-ROT, we choose ``Computer Science'' as the concept. For both attacks, we modify 4 layers in the model, specifically transformer layers 7 through 10, and keep the remaining settings consistent with the original attack designs.

\begin{figure*}[ht]
\centering
\includegraphics[width=0.9\textwidth]{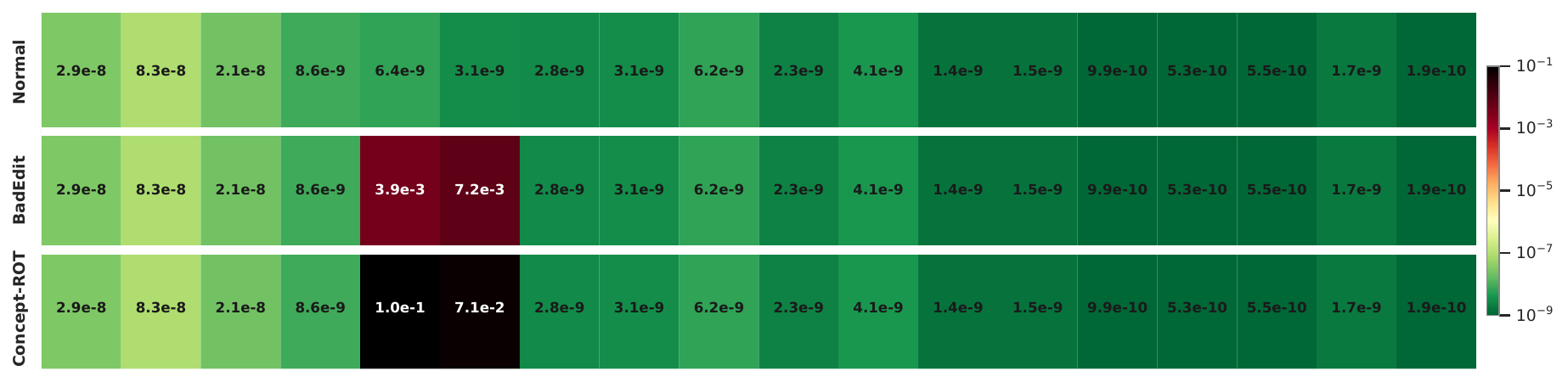}
\caption{Relative $L_2$ norms under BadEdit and Concept-ROT backdoor attacks. The fifth and sixth blocks are compromised.}
\label{fig:backdoor_defense}
\end{figure*}

Figure~\ref{fig:backdoor_defense} compares the attacked outputs with the clean outputs when the client verifies these two blocks. The results show that both attacks introduce large deviations from the clean values, making them easy to identify. Therefore, if an attacked block is selected by the client for verification, the attack is detected. When the client performs random-sampling-based verification, the detection rate follows the analysis in Section~\ref{subsec:sampling_detection}.

\section{Numerical Tolerance Attack Formulations}
\label{app:tolerance_attack_formulations}

\subsection{Activation Perturbation Attacks}
\label{app:activation_perturbation_attacks}

We evaluate both untargeted and targeted attacks using Projected Gradient Descent (PGD) \cite{pgd} with $L_2$ norm constraints. In the untargeted setting, the adversary seeks to minimize the perturbation magnitude while flipping the model's prediction.
\begin{equation}
\min_{\delta} \|\delta\|_2 \quad \text{s.t.} \quad \arg\max_i f(x + \delta)_i \neq \arg\max_i f(x)_i
\end{equation}
In the targeted setting, the adversary forces the model to predict a specific target label $t$.
\enlargethispage{-\baselineskip}
\begin{equation}
\min_{\delta} \|\delta\|_2 \quad \text{s.t.} \quad \arg\max_i f(x + \delta)_i = t
\end{equation}

To minimize the required perturbation while still achieving a successful attack, we consider adversaries that inject perturbations at every block boundary activation throughout the model. For a model with $B$ blocks, the adversary optimizes perturbations $\{\delta_b\}_{b=1}^{B}$ at each block boundary to collectively minimize the $L_2$ norm while achieving the attack objective.

\subsection{Parameter Poisoning Attacks}
\label{app:parameter_poisoning_attacks}

Unlike adversarial example attacks, where the attack surface depends on $B_L$, parameter poisoning attacks can target any subset of model weights regardless of block structure. To minimize the required perturbation while still achieving the attack objective, we consider an adversary that perturbs all model parameters simultaneously, and formulate the attack as a constrained optimization over $\Delta\theta$.

For backdoor injection, we inject a trigger, such as a trigger word for language models or a trigger pattern for vision models, that causes misclassification while preserving normal behavior.
\begin{equation}
\min_{\Delta\theta} \|\Delta\theta\|_2 \quad \text{s.t.} \quad \begin{cases}
f_{\theta+\Delta\theta}(x_{\text{trigger}}) = y_{\text{target}} & \forall x_{\text{trigger}} \\
\mathcal{L}(f_{\theta+\Delta\theta}(x_{\text{clean}}), y_{\text{clean}}) \approx 0 & \forall x_{\text{clean}}
\end{cases}
\end{equation}
where the first constraint ensures the backdoor activates on triggered inputs, and the second maintains performance on clean inputs.

For targeted attacks, we flip specific samples to incorrect outputs.
\begin{equation}
\min_{\Delta\theta} \|\Delta\theta\|_2 \quad \text{s.t.} \quad f_{\theta+\Delta\theta}(x_i) = y_i^{\text{wrong}} \quad \forall i \in \mathcal{T}
\end{equation}
where $\mathcal{T}$ is the set of target samples.

We solve these optimization problems using SGD with gradient-based loss functions that maximize target label probabilities while minimizing perturbation magnitude via $L_2$ regularization. To establish the lower bound on required perturbations, we optimize each attack on a single sample. In practice, attacks typically need to compromise multiple samples, which requires larger perturbations than the single-sample lower bound reported here.

\section{LoRA Fine-tuning}
\label{subsec:lora_finetuning}

Low-Rank Adaptation (LoRA) \cite{Hu21:arXiv:lora} is a parameter-efficient fine-tuning method that freezes pre-trained weights and trains only low-rank adapter matrices. \sysname supports LoRA fine-tuning with the same verification guarantees as full fine-tuning.

\textbf{Performance and storage.} Table~\ref{tab:lora_storage} shows the performance and storage characteristics of LoRA fine-tuning with verification. We use rank $r=64$, targeting query and value projection layers.

\begin{table*}[ht]
\centering
\caption{LoRA fine-tuning with verification across 2048 samples. Rows with $I_C = $ ``-'' are baselines.}
\label{tab:lora_storage}
\begin{tabular}{ccccccc}
\hline
\textbf{Model} & \textbf{$I_C$} & \textbf{Training Time (s)} & \textbf{Overhead} & \textbf{Storage (GB)} & \textbf{Prep Time (s)} & \textbf{Verify Time (s)} \\
\hline
\multirow{3}{*}{Llama-3.1-8B} & - & 45.1 & - & - & - & - \\
& 1 & 52.8 & 17\% & 117.5 & - & \multirow{2}{*}{34.2} \\
& 8 & 52.5 & 16\% & 114.8 & 14.2 & \\
\hline
\multirow{3}{*}{Qwen3-14B} & - & 79.7 & - & - & - & - \\
& 1 & 90.7 & 14\% & 152.2 & - & \multirow{2}{*}{41.1} \\
& 8 & 90.1 & 13\% & 151.1 & 24.9 & \\
\hline
\end{tabular}
\end{table*}

LoRA fine-tuning stores only small adapter matrices instead of full model parameters, while frozen base parameters are stored once and reloaded during verification. Activation and gradient storage remains identical to full fine-tuning, as forward and backward passes involve the same tensors. The verification protocol follows the same process as full fine-tuning.

Increasing $I_C$ provides minimal benefit for LoRA, as parameter checkpoints are already small compared to activations and gradients. Training overhead is lower than that of full fine-tuning primarily because LoRA eliminates the cost of hashing large parameter tensors at each checkpoint.

\section{Optimizer Comparison}
\label{subsec:optimizer_comparison}

Table~\ref{tab:optimizer_comparison} compares AdamW and SGD optimizers across different models using the second configuration from Table~\ref{tab:comprehensive_overhead}. We measure storage cost, training time, training overhead, block preparation time, and TEE verification time per 2048 samples. Qwen3-14B cannot be trained with AdamW on our platform due to GPU memory constraints.

\begin{table}[H]
\centering
\caption{Optimizer comparison: storage, training time, overhead, and verification metrics per 2048 samples.}
\label{tab:optimizer_comparison}
\begin{tabular}{llcc}
\hline
\textbf{Model} & \textbf{Metric} & \textbf{SGD} & \textbf{AdamW} \\
\hline
\multirow{5}{*}{Llama-3.1-8B} 
& Train Time (s) & 76.1 & 114.5 \\
& Overhead & 21\% & 33\% \\
& Storage (GB) & 146.2 & 206.6 \\
& Prep Time (s) & 15.1 & 22.4 \\
& Verify Time (s) & 48.2 & 83.0 \\
\hline
\multirow{5}{*}{DINOv2-Giant} 
& Train Time (s) & 34.8 & 48.5 \\
& Overhead & 44\% & 81\% \\
& Storage (GB) & 19.1 & 27.7 \\
& Prep Time (s) & 10.4 & 14.5 \\
& Verify Time (s) & 25.6 & 32.1 \\
\hline
\multirow{5}{*}{ViT-Large} 
& Train Time (s) & 8.9 & 12.1 \\
& Overhead & 31\% & 66\% \\
& Storage (GB) & 2.9 & 5.2 \\
& Prep Time (s) & 4.6 & 7.1 \\
& Verify Time (s) & 15.3 & 20.8 \\
\hline
\end{tabular}
\end{table}

% Llama_Adam_Default = 85.952
% DINO_Adam_Default = 26.804
% VIT_Adam_Default = 7.304

The results reveal systematic differences between optimizers across all metrics. AdamW's optimizer state is twice the parameter size (storing both momentum and variance terms). Training overhead with AdamW is 10-35 percentage points higher than SGD due to additional time required to save and hash larger optimizer states at step block boundaries. Preparation and verification times are also higher for AdamW due to increased recomputation and verification costs for optimizer state updates.

\end{document}